\documentclass[twocolumn]{aastex62}
\usepackage{graphicx}

\received{04-24-2019}
\revised{06-27-2019}
\accepted{07-26-2019}
\shorttitle{Absence of dark matter in Milky Way dwarf galaxies}
\shortauthors{Hammer et al.}
\begin{document}

% Use the \preprint command to place your local institutional report
% number in the upper righthand corner of the title page in preprint mode.
% Multiple \preprint commands are allowed.
% Use the 'preprintnumbers' class option to override journal defaults
% to display numbers if necessary
%\preprint{}

%Title of paper
\title{On the absence of dark matter in dwarf galaxies surrounding the Milky Way}

\correspondingauthor{Francois Hammer}
\email{francois.hammer@obspm.fr}

\author{Francois Hammer}
\affiliation{GEPI, Observatoire de Paris, Universit\'e PSL, CNRS, Place Jules Janssen 92195, Meudon, France}
\author{Yanbin Yang}
\affiliation{GEPI, Observatoire de Paris, Universit\'e PSL, CNRS, Place Jules Janssen 92195, Meudon, France}
\author{Jianling Wang}
\affiliation{GEPI, Observatoire de Paris, Universit\'e PSL, CNRS, Place Jules Janssen 92195, Meudon, France}
\affiliation{NAOC, Chinese Academy of Sciences, A20 Datun Road, 100012 Beijing, PR China.}
\author{Frederic Arenou}
\affiliation{GEPI, Observatoire de Paris, Universit\'e PSL, CNRS, Place Jules Janssen 92195, Meudon, France}
%\homepage[]{Your web page}
%\thanks{}
\author{Mathieu Puech}
\affiliation{GEPI, Observatoire de Paris, Universit\'e PSL, CNRS, Place Jules Janssen 92195, Meudon, France}
\author{Hector Flores}
\affiliation{GEPI, Observatoire de Paris, Universit\'e PSL, CNRS, Place Jules Janssen 92195, Meudon, France}
\author{Carine Babusiaux}
\affiliation{Universit\'e de Grenoble-Alpes, CNRS, IPAG, F-38000 Grenoble, France }
\affiliation{GEPI, Observatoire de Paris, Universit\'e PSL, CNRS, Place Jules Janssen 92195, Meudon, France}

%Collaboration name if desired (requires use of superscriptaddress
%option in \documentclass). \noaffiliation is required (may also be
%used with the \author command).
%\collaboration can be followed by \email, \homepage, \thanks as well.
%\collaboration{}
%\noaffiliation

%\date{\today}

\begin{abstract}
This paper presents an alternative scenario to explain the observed properties of the Milky Way dwarf Spheroidals (MW dSphs). We show that instead of resulting from large amounts of dark matter (DM), the large velocity dispersions observed along their lines of sight ($\sigma_{los}$) can be entirely accounted for by dynamical heating of DM-free systems resulting from MW tidal shocks. Such a regime is expected if the progenitors of the MW dwarfs are infalling gas-dominated galaxies. In this case, gas lost through ram-pressure leads to a strong decrease of self-gravity, a phase during which stars can radially expand, while leaving a gas-free dSph in which tidal shocks can easily develop. 

The DM content of dSphs is widely derived from the measurement of the dSphs self-gravity acceleration projected along the line of sight. We show that the latter strongly anti-correlates with the dSph distance from the MW, and that it is matched in amplitude by the acceleration caused by MW tidal shocks on DM-free dSphs. If correct, this implies that the MW dSphs would have negligible DM content, putting in question, e.g., their use as targets for DM direct searches, or our understanding of the Local Group mass assembly history. 
Most of the progenitors of the MW dSphs are likely extremely tiny dIrrs, and deeper observations and more accurate modeling are necessary to infer their properties as well as to derive star formation histories of the faintest dSphs.
% that is induced by the DM self-gravity

\end{abstract}


%% From the front matter, we move on to the body of the paper.
%% Sections are demarcated by \section and \subsection, respectively.
%% Observe the use of the LaTeX \label
%% command after the \subsection to give a symbolic KEY to the
%% subsection for cross-referencing in a \ref command.
%% You can use LaTeX's \ref and \label commands to keep track of
%% cross-references to sections, equations, tables, and figures.
%% That way, if you change the order of any elements, LaTeX will
%% automatically renumber them.
%%
%% We recommend that authors also use the natbib \citep
%% and \citet commands to identify citations.  The citations are
%% tied to the reference list via symbolic KEYs. The KEY corresponds
%% to the KEY in the \bibitem in the reference list below. 
\section{Introduction} \label{sec:intro}
The MW halo within $\sim$ 300 kpc is populated by the two Magellanic Clouds and by ten to a million times fewer massive dwarf-spheroidal galaxies (dSphs). Several tens of MW dSphs \citep{McConnachie2012,Munoz2018,Fritz2018} have now been discovered and their number may continue to increase. Furthermore, observations of dSphs revealed  1D, line-of-sight ({\it los}) velocity dispersions ($\sigma_{\rm los}$) that are considerably larger than expectations from their stellar mass, which has led to assume the presence of an additional, dominant component of dark matter (DM). 
The total mass ($M_{\rm J}$), including DM, was estimated by assuming that in each dSph a single, dispersion-supported stellar population is in dynamical equilibrium in the underlying gravitational potential \citep{Walker2009,Wolf2010}. Further assuming spherical symmetry and neglecting the impact of the MW gravitation, the dynamical mass was deduced from dSph star kinematics using the Jeans equation \citep{Walker2009,Wolf2010}. %For decades it was believed that dwarfs inhabiting the MW halo have been captured for a long time, i.e., comparable to the Universe age, and that their high DM content has shielded them against the MW tidal forces. 

%However the Magellanic Clouds orbital motions measured from HST have evidenced that they reach the MW at a first passage \citep{Kallivayalil2013}. They are loosing their gas during their infall, mostly through ram-pressure exerted by the MW halo hot gas, forming the Magellanic Stream \citep{Hammer2015}. A similar process has been suggested to be responsible of the fact that being much less massive, dSphs are gas-free. This is strongly supported by observations showing a strong dichotomy in dwarf gas content, those beyond 300 kpc being gas-rich, the others being gas-free \citep{Grcevich2009,Grcevich2016}. The epoch of infall for most dSphs is thus still debated, and Gaia recently revolutionised our understanding of their orbits, showing that more than half of them have high eccentricities and several with so large apocenters  \cite{Fritz2018} than they are likely at their first passage like the Clouds. This however depends on the MW mass and profile, chosen by \citep{Fritz2018} to be consistent with most MW structural parameters \cite{Bovy2015} including with the most extended rotation curve \citep{Bhattacharjee2014,Huang2016}.

However, \citet{Hammer2018} have shown that dSph Jeans masses ($M_{\rm J}$) and their ratio to luminosity can be derived from the half-light radius ($r_{\rm half}$) and the gravitational acceleration exerted by the MW, which appears to be at odds with the assumption that dSph kinematics are not affected by the MW. Furthermore, it was found that MW tidal shocks provide enough kinetic energy to dSph stars to predict the observed $\sigma_{\rm los}$ with a good accuracy.  This is supported by the fact that MW tidal shocks are exerted along the radial direction, which coincides with the line of sight along which $\sigma_{\rm los}$ are measured. This casts doubts about the validity of the dynamical mass estimate using the Jeans equation, hence on the estimates of the DM amount in dSphs. Why tidal shocks appear to be the dominant effect of the MW gravitational tides remains to be clarified. In the meantime, \citet{Munoz2018} have considerably improved the photometric measurements of most dSphs, improving the accuracy on $r_{\rm half}$ by one order of magnitude.\\
  
Here, we demonstrate from data analysis that, instead of DM, the MW gravitation through tidal shocks can fully account for the dSph kinematics. In Sect.~\ref{Jeans_val} we show that, assuming self-gravity and using the Jeans equation, one can only predict the projected mass density along the line of sight. We then find that this quantity is highly (anti)correlated\footnote{Along the manuscript we have used a Spearman's rank correlation $\rho$ that does not assume any shape for the relationship between variables; the significance of $\rho$ has been tested using  t= $\rho$ $\sqrt{(n-2)/(1-\rho^2)}$, which is distributed approximately as Student's $t$ distribution with {\it n - 2} degrees of freedom under the null hypothesis.} with the MW distance compared to other DM-related quantities (e.g., mass or 3D mass density). In Sect.~\ref{MW_acc} we demonstrate that this is fully explained by the effect of MW tidal shocks on DM-free dSphs. We also characterize the time-scales that warrant the impulse approximation regime necessary for tidal shocks to be efficient.  Sect.~\ref{Discussion} discusses an astronomical scenario in which gas removal by ram-pressure of gas-rich dSph progenitors ensures the predominance of tidal shocks in MW dSphs, then providing dSph properties very similar to that observed. The sample of 21 dSphs with kinematic data and after removing three outliers has been defined in the Appendix~\ref{data} in which scaling relations are analyzed, and which includes two Tables showing the data and the calculated quantities, respectively.  In the following we assume $M_{\rm stellar}/L_V$ ratio of 2.5 for dSphs without young or intermediate age stars, and 1.5 for Carina, Fornax, and Leo I (see, e.g., \cite{Lelli2017} and references therein).
%and assume $M_{\rm stellar}/L_V$ ratio of 2.5 for dSphs without young or intermediate age stars, and 1.5 for Carina, Fornax \& Leo I .

\begin{figure}
\includegraphics[width=3.4in]{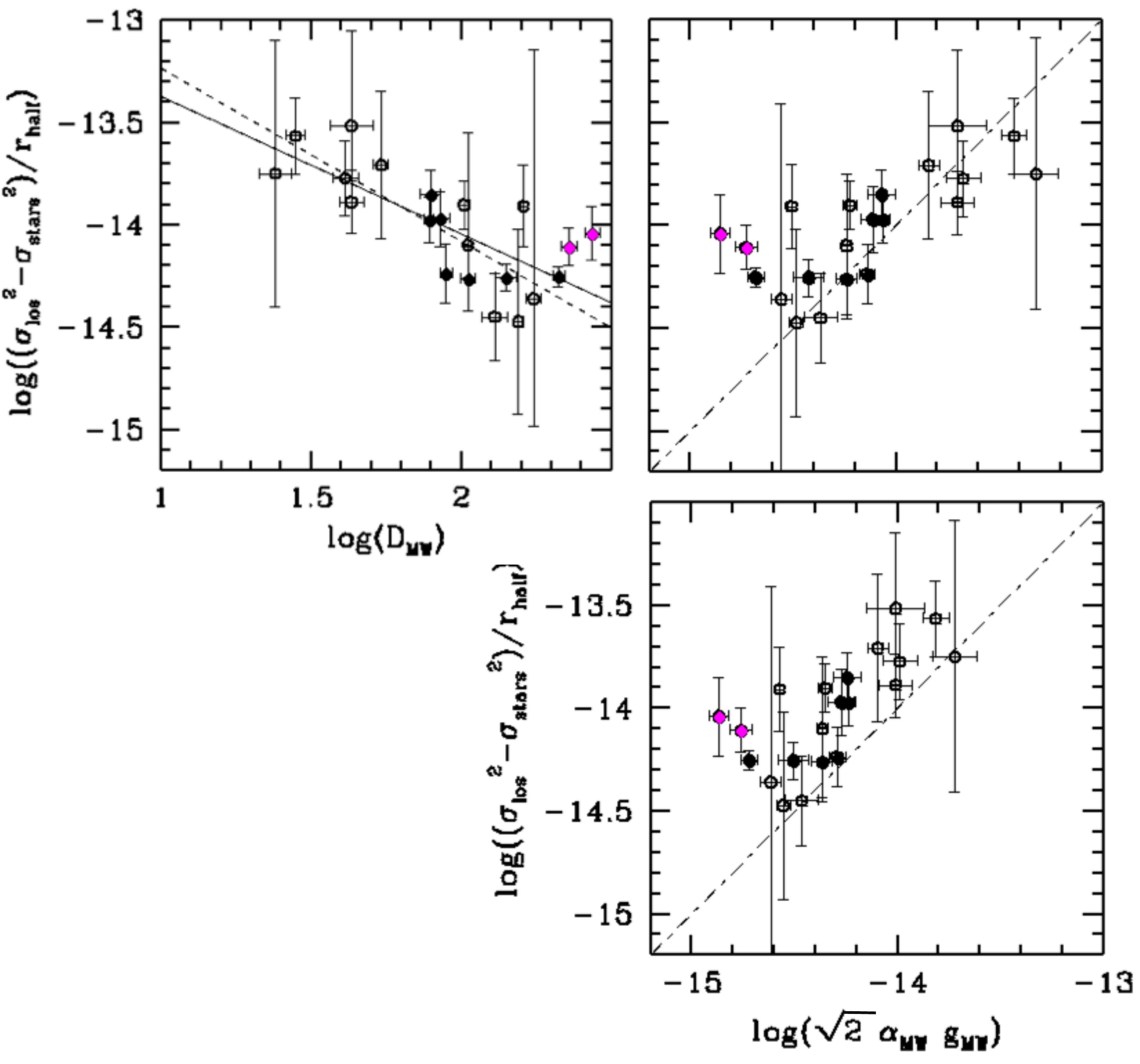}
\caption{{\it Left:} logarithmic anticorrelation between the gravitational acceleration in $km s^{-2}$ that would be due to the DM according to Eq.~\ref{acceleration} and the distance to the MW center in kpc. The best fit for the 21 dSphs is represented by the solid line, while the dashed line represents the best fit after excluding Leo I and Leo II (magenta dots), since they do not obey to the impulse approximation. {\it Right:} same after replacing the MW distance by $\sqrt{2} \times \alpha_{\rm MW} \times g_{\rm MW}$, which is the expected acceleration due to the MW in the frame of the impulse approximation (see Eq.~\ref{MWtides}). The MW mass profile is coming from \cite[top]{Sofue2015} and from \cite[bottom]{Bovy2015}, respectively.}
\label{fig1}
\end{figure}

\section{Limitations on the Use of the Jeans equation to estimate DM}
\label{Jeans_val}
\citet{Wolf2010} derived the dSph total mass within $r_{1/2}$=$4/3 \times r_{\rm half}$  ($r_{1/2}$ and $r_{\rm half}$ are the 3D and 2D half-light radii, respectively, see also \citealt{Walker2009}). Assuming self-equilibrium and using the Jeans equation:
\begin{equation}
M_{\rm J}/M_{\odot} = 930  \times (r_{\rm half}/[1pc]) \times (\sigma_{\rm los} /[1km\times s^{-1}])^2
\label{Wolf}
 \end{equation}
The above mass estimate  (see, e.g., \citealt{Walker2009,Wolf2010}) further assumes that the DM is spherically distributed, although attempts have been made to minimize the impact of the unknown value of the velocity dispersion anisotropy \citep{Wolf2010}. Since the determination of DM is based on a {\it los} velocity dispersion (e.g., $\sigma_{\rm los}$), only the mass density projected along the line of sight can be determined without any assumption on anisotropy\footnote{To illustrate this, let us consider a mass distribution very elongated along the line of sight: the induced $\sigma_{\rm los}$ would indeed be much larger than if the (same) mass was spherically distributed.}. Under the spherical symmetry assumption, it is also a surface mass density ($\Sigma_{\rm J}$) proportional to $M_{\rm J} \times r_{\rm half}^{-2}$.  The {\it los} projected mass density is proportional to the gravitational acceleration caused by the total mass $M_{\rm J}$, whose projection along the line of sight is assumed to induce the observed {\it los} velocity dispersion $\sigma_{\rm los}$.\\
One may try to isolate the effect of the DM in Eq.~\ref{Wolf} by removing the small and known contribution of the stellar mass.
 Within r= $r_{1/2}$, the Jeans mass is the sum of the DM and stellar masses, and the DM surface-mass density is then: 
\begin{equation}
 \Sigma_{DM}= M_{\rm DM} \times r_{\rm half}^{-2} = ( M_J - M_{\rm stellar}/2) \times r_{\rm half}^{-2},\\
  \label{DMdensity}
\end{equation}
In the following, we estimate\footnote{When calculating the 1D velocity dispersion of a Plummer body made of stars only, we cannot assume a constant sigma profile as done in Eq.~\ref{Wolf}; Eq. 31 of \cite{Evans2005} yields $4.80/G \times \sigma_{starsonly}^2  \times  r_{half}$, instead of $4/G \times \sigma_{stars}^2  \times  r_{half}$ in Eq. 2 of \cite{Wolf2010}. It leads to a value of $\sigma_{starsonly}$=  0.9127 $\times$ $\sigma_{stars}$.} the 1D {\it los} velocity dispersion ($\sigma_{\rm stars}$) associated with the sole stellar mass by inverting Eq.~\ref{Wolf} and replacing $M_{\rm J} (r_{1/2})$ by $M_{\rm stellar}/2$. By multiplying Eq.~\ref{DMdensity} by the gravitational constant {\it G}  it leads to the acceleration caused by the DM ($a_{\rm DM}= G M_{\rm DM} \times r_{\rm half}^{-2}$) for which only its projection along the line of sight is robustly known, which is:
%\footnote{Passing from Eq.~\ref{Wolf} to Eq.~\ref{acceleration} assumes only the conservation of the masses within r=$r_{1/2}$ providing the DM role by subtracting to the Jeans mass $M_{\rm J}$ a second order quantity, the stellar mass.}:
\begin{equation}
a_{\rm DM}= G M_{\rm DM} \times r_{\rm half}^{-2} = ( \sigma_{\rm los}^2 - \sigma_{\rm stars}^2) \times r_{\rm half}^{-1}.
  \label{acceleration}
\end{equation}

Left panel of Figure~\ref{fig1} shows that the acceleration due to the DM is strongly anticorrelated ($\rho= 0.76$ and $t= 5.1$, see Appendix~\ref{stats}) with the distance to the center of the MW ($D_{\rm MW}$). Could this be due to a combination of correlations, e.g., between surface mass density ($\Sigma_{\rm DM}$) and radius, and between radius and MW distance? Appendix~\ref{correlations} indicates that $\sigma_{\rm los}$ correlates well ($\rho$=0.80 for the 21 dSph sample) with $r_{\rm half}$, following:
\begin{equation}
\log(\sigma_{\rm los})=0.37 \times \log(r_{\rm half})+0.028 
 \label{sig_r}
\end{equation}
Combining Eq.~\ref{sig_r} with Eq.~\ref{Wolf}, it results that $M_{\rm J}$ scales as $r_{\rm half}^{1.75}$. If spherical symmetry is assumed, $\Sigma_{DM}$= $M_{\rm J} \times r_{\rm half}^{-2}$ should depend weakly on  radius, i.e., $\propto$ $r_{\rm half}^{-0.25}$. Such a behavior is verified by the data (logarithmic slope=-0.33 with $\rho$= 0.49). The radius is weakly correlated with the distance (see Appendix~\ref{correlationsDMW}, logarithmic slope=0.87 with $\rho$= 0.45). Combining a weak anticorrelation with a weak correlation can neither provide the strong anticorrelation nor the slope (-0.62) of the relation shown in the left panel of Figure~\ref{fig1}. \\
Here we have discovered that the most accurately determined DM quantity, $\Sigma_{\rm DM}$ (or $a_{\rm DM}$), is unexpectedly anticorrelated to the MW distance. This indicates that MW dSphs cannot be further considered as isolated, which questions the validity of the DM mass estimates provided by self equilibrium deduced from the Jeans equation.
%If the DM was entirely responsible This indicates that MW dSphs cannot be further considered as isolated and it questions the mass estimate provided by the Jeans equation.
%If the DM was entirely responsible for the $\sigma_{\rm los}$ enhancement through the Jeans equation, the projected DM mass density (or the associated gravitational acceleration) would only depend on dSph intrinsic properties, but not on a property linked to the host galaxy, disproving further that dSphs can be considered as isolated.\\ %This invalidates Eq.~\ref{Wolf} for estimating the mass. \\

\begin{figure*}
\includegraphics[width=6.2in]{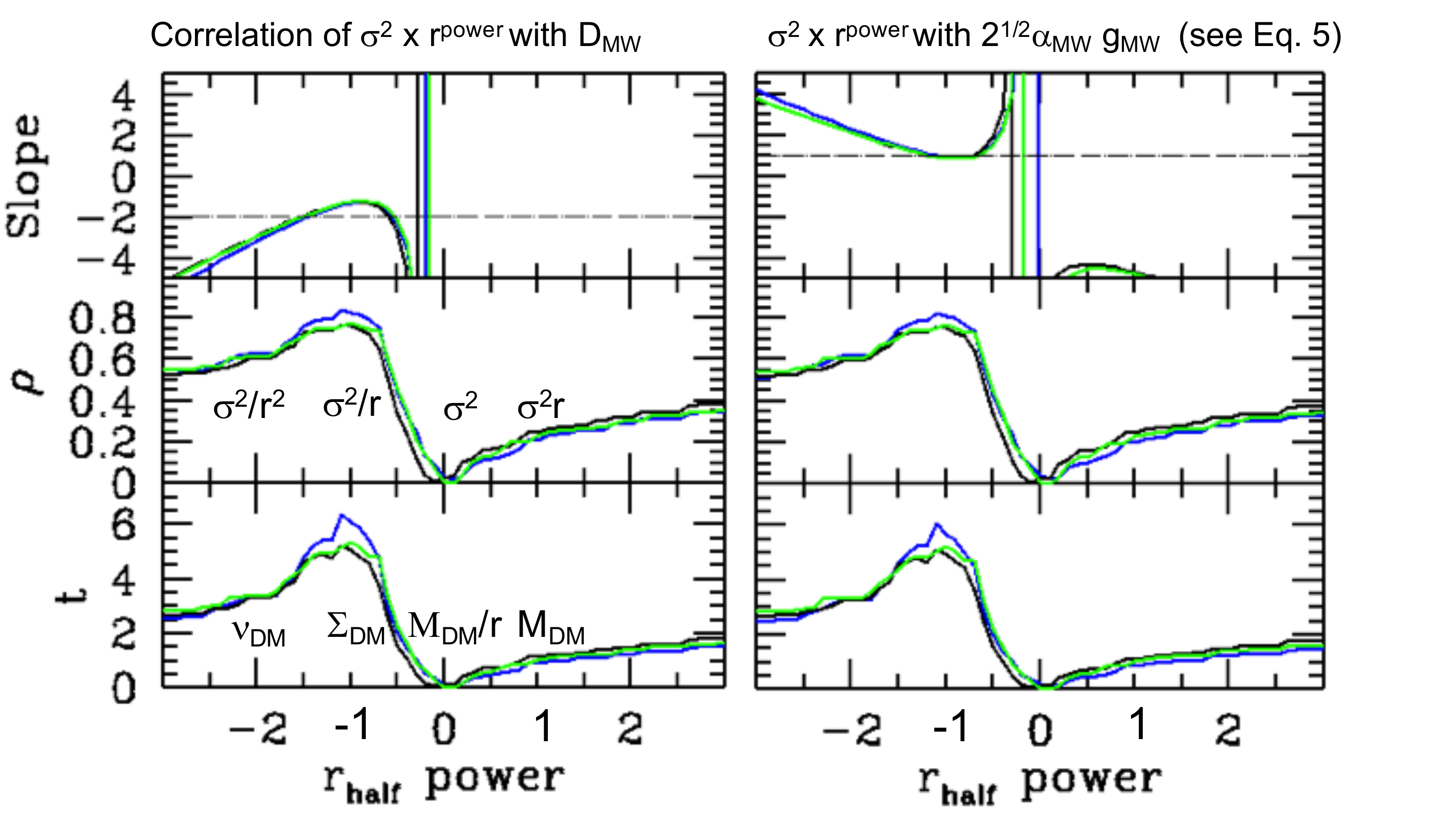}
\caption{Summary of logarithmic correlations (or absence of correlations) between dynamical quantities and $D_{MW}$ (left) and $\sqrt{2} \times \alpha_{\rm MW} \times g_{\rm MW}$ (right), respectively. From top to bottom, the figure shows the slope, correlation strength, and Student $t$ parameters when Eq.~\ref{acceleration} is replaced by $(\sigma_{\rm los}^2 - \sigma_{\rm stars}^2) \times r_{\rm half}^{\rm power}$ (dubbed $\sigma^2 r^{\rm power}$). {\it Left:} It shows how the correlation changes with the power index on $r_{\rm half}$. Observed physical quantities are indicated in the middle panel  (right side) of Eq.~\ref{acceleration}. Calculations of DM-related quantities require the assumption of isotropy and absence of significant effects from MW tides (see the text), and they are indicated in the bottom panel. They include the mass and its 1D, 2D, and 3D density, $M_{\rm DM}$, $M_{\rm DM}/r$, $\Sigma_{\rm DM}$ and $\nu_{\rm DM}$ respectively. Correlation strength and $t$ peak at power = -1, which represents the DM {\it los} acceleration (see Eq.~\ref{acceleration}). Black and green curves are based on the dSph galaxy sample of 21 dwarfs defined in the Appendix~\ref{data}, while the blue curve is for the same sample but Leo I and Leo II.  The green curve illustrates a case for which the $M_{\rm stellar}/L_V$ ratios are multiplied by 2. {\it Right:} same but for the correlation with $\sqrt{2} \times \alpha_{\rm MW} \times g_{\rm MW}$, for which we use the MW model from \cite{Bovy2015}. Using the MW model from \cite{Sofue2015} provides very similar graphs.}
\label{fig2}
\end{figure*}

The left side of Figure~\ref{fig2} is made up of three panels showing the slope, correlation strength, and Student $t$ parameter of a generalized correlation between $( \sigma_{\rm los}^2 - \sigma_{\rm stars}^2) \times r_{\rm half}^{\rm power}$ and $D_{\rm MW}$. By replacing $r_{\rm half}^{-1}$ by $r_{\rm half}^{\rm power}$ in Eq.~\ref{acceleration}, it identifies the DM mass and its 1D to 3D-mass densities  (see indications in the bottom panel), after neglecting the effect of MW tides.  As for Eq.~\ref{Wolf}, the above equations assume spherical symmetry for the dSphs and DM-related quantities are calculated at r= $r_{\rm 1/2}$= 4/3 $\times$ $r_{\rm half}$.\\

The right side of Eq.~\ref{acceleration} illustrates dSph observed quantities ($\sigma_{\rm los}$, $r_{\rm half}$, $M_{\rm stellar}$) without any dependence on modeling or on the nature of dSphs, and Figure~\ref{fig2} is only based on them. Concerning the DM content of dSphs,  Figure~\ref{fig2} is illustrative because:
\begin{itemize}
\item Data show significant anticorrelations for power $\le$ -1, with a prominent peak at {\it r} power= -1 ($\Sigma_{\rm DM}$), i.e., for the quantity robustly estimated from observations; it suggests that the anticorrelation between the DM 3D-mass density, $\nu_{DM}$, and the MW distance is only due to that between $\Sigma_{\rm DM}$ with $D_{\rm MW}$;   
\item Data confirm the absence of correlation between $\sigma_{\rm los}$ and distance ({\it r} power = 0), as well as between the assumed total DM mass ({\it r} power = 1) derived from Eq.~\ref{Wolf} and the MW distance.
\end{itemize}
%The above has a considerable impact since it is based only from dSph observable quantities ($\sigma_{\rm los}$, $r_{\rm half}$, $M_{\rm stellar}$) without any dependence on modeling nor on the nature of dSphs. 
This may also explain why the ``DM in isolated dSph'' hypothesis received so much credit, given that its associated mass appears to not be correlated with the host galaxy properties. Reproducing Figures~\ref{fig1} and ~\ref{fig2} is likely a challenge for the standard scenario of DM-dominated MW dSphs. For example, why by multiplying $\Sigma_{\rm DM}$ (or $a_{\rm DM}$) by $r_{\rm half}$, which is weakly correlated with the MW distance, the strong anticorrelation found in Fig.~\ref{fig1} disappears (at {\it r} power= 0 in Fig.~\ref{fig2})? \\ 
In fact an anticorrelation between DM 3D-mass density ($\nu_{\rm DM}$) and distance to the MW has been found by \cite{Diemand2008} and by \cite{Moline2017}, in the frame of a tidal stripping of pure DM subhalos, which would be denser when lying near the MW center. Could such an anticorrelation be at the root of the relation shown in Figure~\ref{fig1}? Top left panel of Figure~\ref{fig2} reveals that $\nu_{\rm DM}$ shows a logarithmic slope of -2.9, implying that at 30 kpc from the MW, sub-haloes would be 340 times denser than at 250 kpc. However, stellar densities stay unchanged within the same MW distance range, and it would require a considerably more efficient stripping of DM\footnote{If DM particles have strongly radially biased orbits (but see \citealt{Taylor2011}) they will be more affected by tidal stripping than stars.} than that of stars while they lie within a similar volume in the dSph (see details in Appendix~\ref{DMChallenge}).

%However this neglects baryons and the fact that MW also acts through ram-pressure effects in removing the gas, which leads instead to a dominance of tidal shocks \cite{Mayer2007,Mayer2011,Kazantzidis2017}. The challenge for the standard scenario would be to reproduce together, the correlations and absence of correlations described in Figure~\ref{fig2} as well as being consistent with the scaling relation shown in the Supplemental Material, while providing realistic dSphs comparable to the observations.}

\section{Why is the acceleration correlating with the MW distance?}
\label{MW_acc}
Indeed the anticorrelation between $( \sigma_{\rm los}^2 - \sigma_{\rm stars}^2) \times r_{\rm half}^{-1}$ and $D_{\rm MW}$ hints to Eqs. (2) and (B16) of \cite{Hammer2018} established in the frame of the impulse and distant-tide approximations, and assuming DM-free dSphs. They show that MW tidal shocks bring sufficient kinetic energy to generate an additional term to the {\it los} velocity dispersion, which is $\sigma_{\rm MW}=\sqrt{\sigma_{\rm los}^2 - \sigma_{\rm stars}^2}$. Note that both accelerations, which are due to MW tidal shocks on DM-free dSphs and that are assumed to be due to DM in the frame of self-gravity, are indeed projected along the MW radial direction that coincides with the line of sight. Eq. B16 of \cite{Hammer2018} can be written as:
 \begin{equation}
\sigma_{\rm MW}^2= \sigma_{\rm los}^2 - \sigma_{\rm stars}^2=  \sqrt{2} \: \alpha_{\rm MW}\;  g_{\rm MW} \: r_{\rm half} \: 
 \label{MWtides} 
\end{equation}
 where $g_{\rm MW}$= $GM_{\rm MW}/D_{\rm MW}^{2}$ is the gravitational acceleration of the MW and
  \begin{equation}
 \alpha_{\rm MW} = 1 - \partial log(M_{\rm MW})/\partial log(D_{\rm MW}) 
  \label{alpha} 
\end{equation}
  characterizes the shape of the MW mass profile, taking values close to 1 in the Galactic outskirts ($\partial$$M_{\rm MW}$/$\partial$$D_{\rm MW}$$\sim$ 0) and close to 0 when the slope of the MW mass profile is steep. \\
  
Can Eq.~\ref{MWtides} be at the root of Figure~\ref{fig1}? To verify this we have to compare $( \sigma_{\rm los}^2 - \sigma_{\rm stars}^2) \times r_{\rm half}^{-1}$ with $\sqrt{2} \times \alpha_{\rm MW} \times  g_{\rm MW} $. The later term requires the knowledge of the MW potential and we have considered two models reproducing the MW kinematics \citep{Bovy2015} and extended rotation curve \citep{Sofue2015} from observations of distant massive stars and globular clusters. Right panels of Figure~\ref{fig1} show that the correlation between the two quantities is as strong as the one shown in the left panel. Moreover, for most dSphs, within their error bars, the {\it los} gravitational acceleration attributed to the DM self-gravity equals that caused by MW tidal shocks on DM-free dSphs. Why would the acceleration caused by the DM be precisely what it is expected from MW tidal shocks on DM-free dSphs? {\bf In other words, why does the latter predict that the DM mass-to-light ratio of Segue 1 is several thousand, while that of Fornax is only 10?}

\begin{figure}
\includegraphics[width=3.4in]{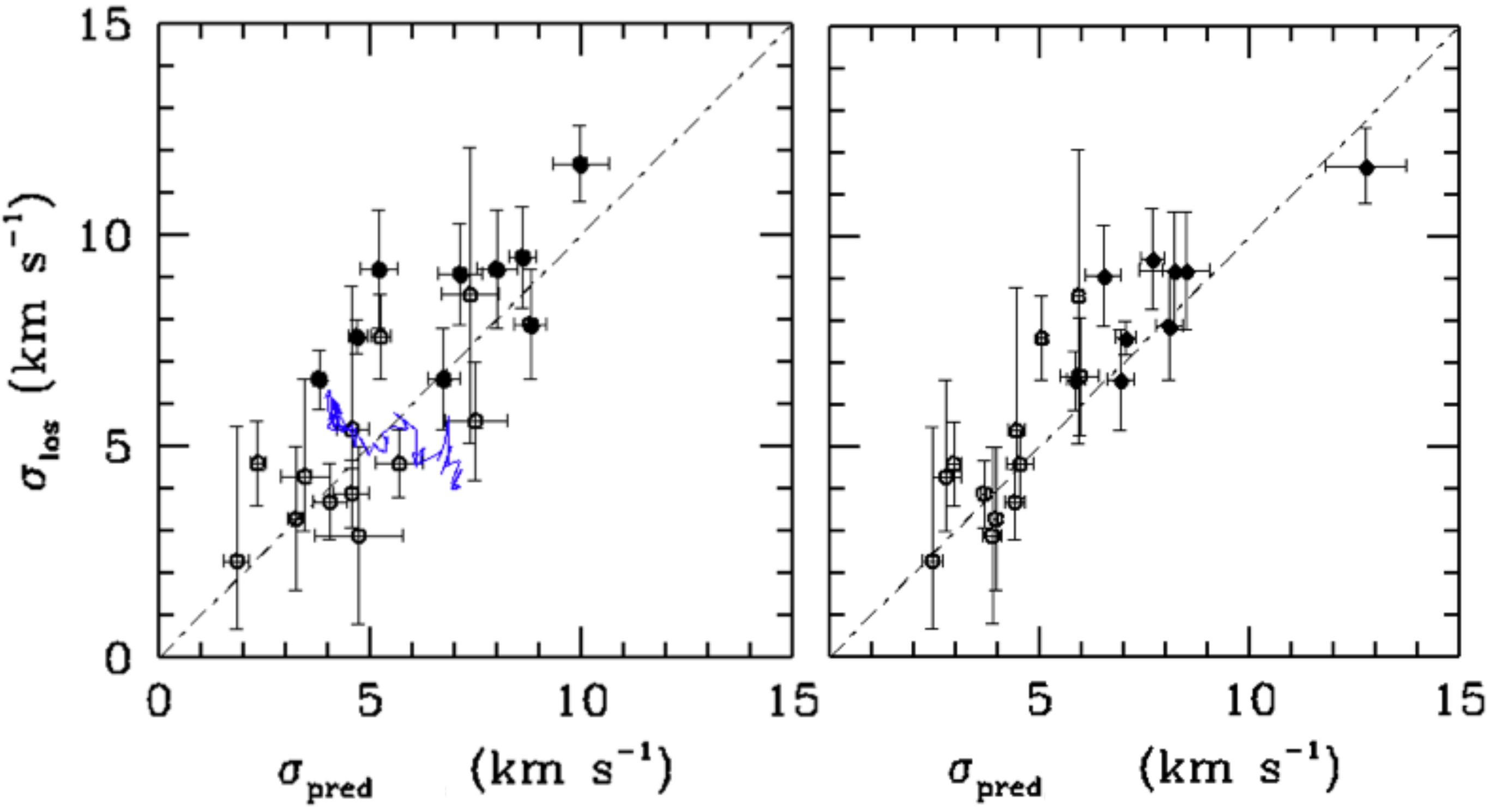}
\caption{Correlation between the observed $\sigma_{\rm los}$ and that predicted from Eq.~\ref{MWtides}, i.e.,  $\sigma_{pred}$ = $\sqrt{\sigma_{\rm MW}^2 + \sigma_{\rm stars}^2}$. {\it Left:} It assumes the MW mass and its profile from \cite{Sofue2015}. The blue line indicates the track of a simulated dwarf falling into the MW halo, adapted from \citet{Yang2014} by adopting a softening radius of 0.5 pc, i.e., warranting that almost no stellar particles are affected by non-Newtonian motions (see Figure~\ref{movie}). {\it Right:} same after optimizing the MW mass profile.} % to reach a least square value of 1.02. }
\label{fig3}
\end{figure}

%Here we verify whether or not observed star motions in dSphs can be explained by MW tidal shocks acting on DM-free dSphs.
Figure~\ref{fig3} shows the comparison between predictions from Eq.~\ref{MWtides} to the observed {\it los} velocity dispersions ($\sigma_{\rm los}$) in dSphs. In the left panel, MW model from \cite{Sofue2015} is found to predict with good precision the observations, with slope = 0.9 and a correlation coefficient $\rho = 0.78$. Similar values (slope = 1.17 and  $\rho = 0.89$) are found when using the MW model from \cite{Bovy2015}. %The probability that this is just a coincidence are 1.5 $10^{-5}$ and 3 $10^{-8}$, respectively. 
We notice that the two models are based on the \citet{Navarro1997} density profile that does not converge at large distances. It indeed excludes values of $\alpha_{\rm MW}$ larger than 0.7, which allows significant room to use more appropriate MW profiles and optimize them. Although it is beyond the scope of this paper to derive the MW total mass and its profile, the right panel of Figure~\ref{fig3} illustrates the result of a very first attempt, which leads to a total MW mass and a mass profile in very good agreement with mass and extended rotation curve profiles from \cite{Bhattacharjee2014} and \cite{Huang2016}. \\

The right panel of Figure~\ref{fig2} illustrates the correlation between $(\sigma_{\rm los}^2 - \sigma_{\rm stars}^2) \times r_{\rm half}^{\rm power}$ and $\sqrt{2}$  $\alpha_{\rm MW}\  g_{\rm MW}$, the former term involving only observed properties, the later term involving only the MW gravitation and mass density profile. It is essentially similar to the left panel, although the slope of the correlation is exactly 1 for power = -1, which indicates that Eq.~\ref{MWtides} fully accounts for energy exchanges between the MW and the dSphs along the line of sight, and that within an acceleration range of almost 2 decades.

  Eq.~\ref{MWtides} applies to DM-devoid dSphs orbiting into the MW halo, and it is based on the fact that the sizes of the dSphs are much smaller than their distances to the MW (the distant-tide approximation) allowing one to account for smooth variations of the MW potential across dSphs (see Eq. 8.34 in \citealt{Binney2008}). Furthermore, Eq.~\ref{MWtides} requires that dSph stars undergo a fluctuation of the MW gravitational potential over time scales that are smaller than their present-day crossing-time scales ($t_{\rm cross}$= $r_{\rm half}/\sigma_{\rm los}$). It has been demonstrated by \citet{Aguilar1985} that such an impulse approximation holds even for slow encounters (or those with rather low-eccentricities), i.e., if the encounter is as long as, or a few times longer, than $t_{\rm cross}$ (see also \citealt{Binney1987,Binney2008}). \\
 
 This situation might seem similar to that of globular clusters (GCs) progressively destroyed by tidal shocks during their passages near the MW bulge \citep{Aguilar1993,Gnedin1999a,Gnedin1999b}. However, dSphs have central densities 10,000--100,000 times smaller than GCs, and their interaction with the MW potential is much more destructive during a single dynamical time $t_{\rm dyn}$= $r_{\rm half}/\sigma_{\rm starsonly}$ (see footnote 3), which ranges from $5\times10^{7}$ to $9\times10^{8}$ yr (see the second table of Appendix~\ref{data}). \citet{Gnedin1999b} showed that it takes more than 3 (resp. 10--15) dynamical times for a system to virialize (resp. relax)  after a perturbation. In the course of their long orbits\footnote{For example, with $t_{\rm orbit}$$\sim$ $D_{\rm MW}/(2 \times V_{\rm GSR}$), where $V_{\rm GSR}$ is the galactocentric projection of the radial velocity. Replacing $V_{\rm GSR}$ by $V_{\rm rad}$ calculated by \citet{Fritz2018} does not change the results.} within the halo, dSphs are affected by changes of the energy brought by the MW tides. The encountering time ($t_{\rm enc}$) can be evaluated by assuming that these changes have to be larger than the 
 kinetic energy associated with the self-gravity of the DM-free dSphs ($1/2 \times \sigma_{\rm starsonly}^{2}$).  Differentiating Eq.~\ref{MWtides}  yields:
  \begin{equation}
  \Delta(\sigma_{\rm MW}^{2})/\Delta(D_{\rm MW}) =  -2\sqrt{2} \: \alpha_{\rm MW}\;  \times GM_{\rm MW}/D_{\rm MW}^{3} \: r_{\rm half},
  \label{DD_MW_1} 
\end{equation}
in which we have neglected the (slow) variation of $\alpha_{\rm MW}$ with $D_{\rm MW}$. Then:
  \begin{equation}
 \Delta(\sigma_{\rm MW}^{2}) = -2 \times \sigma_{\rm MW}^{2} \times  \Delta(D_{\rm MW})/D_{\rm MW},
 \label{DD_MW_2} 
\end{equation}
  
  hence a distance change of $\Delta(D_{\rm MW})$ induces a change of $\Delta(\sigma_{\rm MW}^{2})$ equal to twice the kinetic energy due to DM-free dSph self-gravity (1/2$\sigma_{\rm starsonly}^{2}$), if:
  \begin{equation}
 \Delta(D_{\rm MW}) =  \sigma_{\rm starsonly}^2/\sigma_{\rm MW}^{2} \times D_{\rm MW}/2
 \label{DD_MW} 
\end{equation}
%Table~\ref{tab2} (see Appendix) shows that
Given the small dSph stellar masses, the ratio $\sigma_{\rm starsonly}^{2}/\sigma_{\rm MW}^{2}$ is usually so small (see Appendix~\ref{data}) that changing $D_{\rm MW}$ by only a few percent warrants a kinetic energy change larger than $1/2 \times \sigma_{\rm starsonly}^{2}$. In the impulse approximation, the perturbation causes an instantaneous change in the velocity of each star that depends only on its position \citep{Aguilar1985}. It holds when $t_{\rm cross}$ is larger than the encountering time $t_{\rm enc}$=$\Delta(D_{\rm MW})$/$V_{\rm GSR}$, and eventually leads to Eq.~\ref{MWtides}. Another way to establish the validity of Eq.~\ref{MWtides} is to consider the stars that are in resonance with the MW gravitational perturbation: following \cite{Weinberg1994} and \cite{Gnedin1999a} the fraction of resonant stars follows $(t_{\rm enc}/t_{\rm cross})^{-1}$, which is 100\% for most dSphs.
Values of $t_{\rm enc}/t_{\rm cross}$ are below 1 for most dSphs, but larger than 10 for Leo I and Leo II (see Appendix~\ref{data}). Intermediate values for Carina and Fornax are likely yielding results consistent with the impulse approximation even if the conditions are not strictly satisfied \citep{Aguilar1985,Binney1987}. Moreover, stars in these dSphs would not have time to come back to equilibrium (virialization), since the ratio between $t_{\rm enc}$ and the virializing time (3 $\times t_{\rm dyn}$) is always below or very close to 1, except for Leo I and Leo II. It is noteworthy to see that Leo I and Leo II are also outliers in the relations shown in Figure~\ref{fig1} (see magenta dots). 

{\bf
\section{Tidal shocks May explain the MW dSph kinematics}
}
\label{Discussion}
It is widely accepted that dSph progenitors have probably lost their gas through ram-pressure effects caused by the diffuse MW halo hot gas \citep{Mayer2001,Grcevich2009}. This is supported by a strong dichotomy in dwarf gas content, those beyond 300 kpc being gas-rich, the others being gas-free \citep{Grcevich2009}. The old stellar ages and low metallicities of most of the dSphs have been often considered as evidence for an ancient infall of primordial dwarf galaxies. However, there is mounting evidence for a more recent infall of most dSphs. 74\% of them \citep{Fritz2018} have orbital motions along a gigantic plane perpendicular to the MW disk suggesting a common infall \citep{Pawlowski2014} with the Magellanic Clouds (MCs), which also belong to this gigantic plane and are at their first passage \citep{Kallivayalil2013}. The MCs show large orbital eccentricities, a property shared by most dSphs as inferred from {\it Gaia} measurements \citep{Fritz2018}, with half of them having apocenters in excess of 300 kpc. This is especially true for MW mass\footnote{\citet{Fritz2018} have also considered another mass model, which is like that of \citet{Bovy2015} but for which they multiply the halo DM mass by a factor 2 (their $M_{MW}= 1.6 \times 10^{12} M_{\odot}$ model). It is however discrepant with the extended rotation curve of the MW.} and profile consistent with most MW structural parameters \citep{Bovy2015}. In the following we show how the above astronomical context leading to Eq.~\ref{MWtides} can fully account for the large measured values of dSph velocity dispersions.\\

%However, the epoch of infall for most dSphs is still debated, and Gaia recently revolutionized our understanding of their orbits, showing that more than half of them have high eccentricities, and several with large apocenters \cite{Fritz2018}. Moreover 74\% of them \cite{Fritz2018} have orbital motions in a gigantic plane perpendicular to the MW disk suggesting a common infall \cite{Pawlowski2014}, including with the Magellanic Clouds (MCs), which are at their first passage \cite{Kallivayalil2013}. %while loosing their gas through ram-pressure exerted by the MW halo hot gas, forming the Magellanic Stream \cite{Hammer2015}. 
%The MCs also show large orbital eccentricities, a property shared by most dSphs from their Gaia orbits \cite{Fritz2018}, and half of them have apocenters in excess to 300 kpc. This is especially true for MW mass and profile, which is consistent with most MW structural parameters \cite{Bovy2015}. It suggests further a common infall with the Clouds and relatively recent passage for many dSphs. In the following we show that it is the above astronomical context that leads to Eq.~\ref{MWtides}, which fully explains the large values of dSph velocity dispersions.\\

Eq.~\ref{MWtides} has been established under the impulse approximation regime (tidal shocks), which is robust for all dSphs but Leo I and II. The factor $\sqrt{2}$ in Eq.~\ref{MWtides} results from the assumed Plummer model for the stellar density in dSphs, assuming further a spherical symmetry and using a radius that is the geometric mean between the major and minor half-light radii. Adopting different models does not affect Eq.~\ref{MWtides} by more than a few percent \citep{Hammer2018}. \\
However, Eq.~\ref{MWtides} does not account for tidal stripping. The global instantaneous energy change $\Delta E$ caused by the MW tides on an individual star with velocity $\bf v$ is:
\begin{equation}
 \Delta E = {\bf v \cdot \Delta v} + 1/2 (\Delta v)^2
\label{fullenergy}
\end{equation}
where ${\bf \Delta v}$ is the change in velocity (see Eq. 7--41 of \citealt{Binney1987}). If dSphs are spherically symmetric, the first term (called "tidal stripping" or "diffusion term," see \citealt{Binney2008}, P. 663) vanishes when averaged over all stars letting the second term (called "tidal shocking" or "heating," here approximated to $1/2 (\Delta \sigma)^2$) lead to Eq.~\ref{MWtides} \citep{Hammer2018}. The first comprehensive simulations of DM-devoid dwarfs falling into the gravitational potential of the MW has been done by \citet{Piatek1995}, \citet{Kroupa1997}, and \citet{Klessen1998}, all assuming gas-free progenitors. During their approach, progenitors are initially at large distances from the MW, which results in weak gravitational field and $\mid\Delta v\mid \ll\mid v\mid$, leading to a dominant first term in Eq.~\ref{fullenergy} as described by the diffusive regime (see \citealt{Binney2008}, P. 663). This led \cite{Piatek1995} to model dSphs that experience large star losses by tidal stripping. However, these modeled dSphs were too elongated along the line of sight and had kinematics that were not consistent with observations (see also \citealt{Klessen2003,Read2006}). Conversely, in MW dSphs, $\mid\Delta v\mid$ (approximated by $\sigma_{\rm MW}$) is systematically larger than $\mid v\mid$, if approximated by $\sigma_{\rm stars}$. Furthermore, observations of RR Lyrae stars reveal that dSphs are not particularly elongated along the line of sight \citep{Hernitschek2018}, consistently with expectations and simulations in which tidal shocks dominate. \\
%Conversely, when tidal stripping dominates, simulations of DM and gas-free progenitors ,Read2006} provide dSphs that are too elongated along the line of sight \citep{Klessen2003,Read2006}.
The above considerations point out a mechanism that has prevented significant tidal stripping of stars during the infall. In fact, if gas dominates dSph progenitors as expected from observations \citep{Grcevich2009}, it would have been easily removed by ram pressure stripping.  % and which would have shielded them against the diffuse regime. 
Gas removal leads to a serious decrease of self-gravity, a phase during which stars are left almost free to expand within a spherical geometry, preventing the diffuse regime and letting tidal shocks dominate. \\

\begin{figure*}
\includegraphics[width=6in]{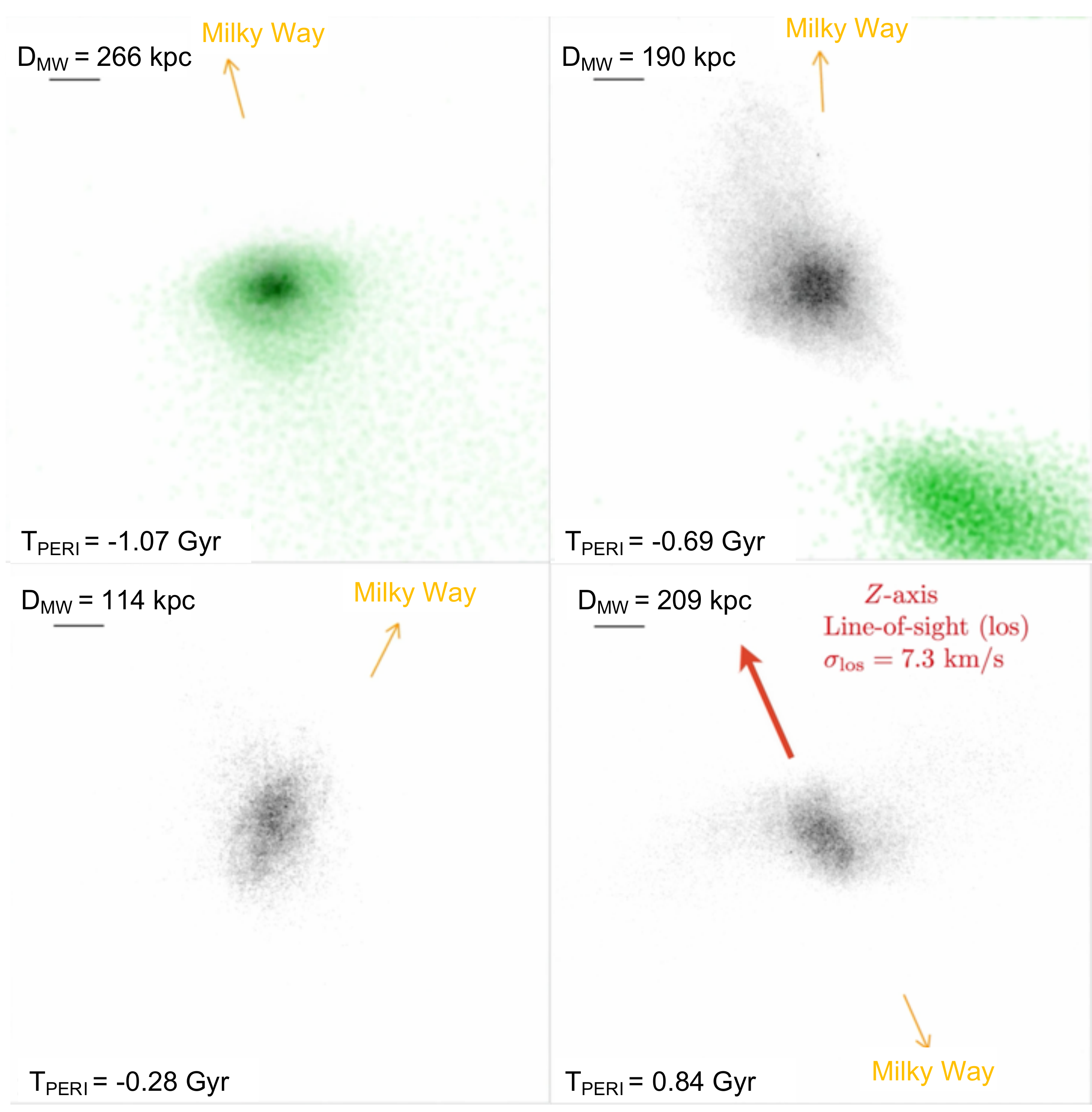}
\caption{Transformation of a DM-devoid and gas-rich dwarf into a dSph after a first passage into the MW halo.  An animation is available that shows a close-up view of the dwarf galaxy in the orbital plane that includes the MW;  distance to the dwarf and direction (see orange arrows) are indicated in the top left of each panel. The initially gas-rich dwarf galaxy (H I gas in green, stars in black) infalls into the MW halo, where hot gas content exerts ram-pressure onto the HI gas content as seen in the top-left panel where the dwarf motion is almost along the direction of the MW (orange arrow). The top-right panel is a snapshot taken after the HI gas has been fully removed from the dwarf, and since the gas accounted for 71\% of the total initial mass, it left the stars free to expand over all directions due to the decrease of self-gravity. The bottom left panel shows the dwarf closely approaching its pericenter (time to pericenter is indicated in the bottom left of each panel) and expanded stars are mostly affected by MW tidal shocks that let them elongate along the line-of-sight direction that matches the direction of the MW gravitational force. The bottom right panel shows the dwarf after pericenter, at which time stars are continuously elongated along the line of sight, providing the large observed values of $\sigma_{\rm los}$ (see the value and red arrow).
This simulation is adapted from one of the simulations published in Yang et al. (2014), i.e., the model $TDG3-rp28$ in their Table 3. It has been improved by adopting a gravitational force softening of 0.5 pc instead of 50 pc for stellar particles (17 $M_{\odot}$/particle, 500,000 particles) in dwarf galaxies. Given the stellar density in dSphs, there should be no or very few stellar particles that can be affected by the undesirable effects due to the softening. These simulations are based on the hydrodynamical / N-body code Gadget2 (Springel 2005), and use simulated DM-free galaxies with initial mass and gas fraction (within a 3 kpc projected radius) of 1.35$\times 10^8$ $M_{\odot}$ and 71\%, respectively. The simulation in the video begins at T$_{\rm PERI}$ = -1.635 Gyr and ends at 0.839 Gyr. The video duration is 71 s.
}
\label{movie}
\end{figure*}

Figure~\ref{movie} illustrates the overall process from the gas removal by ram-pressure to the expansion of the stars, and later on, how tidal shocks exerted by the MW dynamically heat the stars along the MW direction, which coincides with the line of sight.  Figure~\ref{fig3} (see also Fig. 5 of \citealt{Hammer2018}) shows how hydrodynamical simulations from \cite{Yang2014} based on gas removal by ram-pressure and then tides from MW can reproduce expectations from Eq.~\ref{MWtides}. These can reproduce the morphologies (see Figs. 9 and 10 of \citealt{Hammer2018}), the stellar density (see Appendix~\ref{correlations}), and the flat velocity dispersion profiles along the dSph radii (see Figs. 5 and 6 of \citealt{Yang2014}, see also calculations in \citealt{Hammer2018}).
The prominence of tidal shocks in the case of gas-rich progenitors has been shown by \cite{Kazantzidis2017}, who also found that most of the very elongated shapes or bar-like features obtained in gas-free simulations \citep{Lokas2012} then vanish.\\

Properties of 'tidally-shocked,' DM-devoid dSphs are likely undistinguishable from those observed. Tidal features are then not expected to be observed in DM-devoid dSph at recent infall, because MW tidal shocking dominates tidal stripping. This is consistent with the two following observations: (1) there are no obvious tidal features in the relatively bright, so-called classical dSphs except perhaps for Carina \citep{Battaglia2012}, which is likely near its pericenter \citep{Fritz2018}; and (2) wide-field observations of several dSphs have revealed stars beyond their DM tidal radius (Carina:  \citealt{McMonigal2014}; Draco: \citealt{Kleyna2001,Segall2007}, Fornax: \citealt{Battaglia2006}).
Conversely to \citet{Aguilar1985}, \citet{Gnedin1999a}, and \citet{Gnedin1999b}, we have not examined the theoretical evolution with time of the DM-devoid dSphs, although our modeling indicates also mass losses, in particular, for the smallest ones. Tidal shocking could lead to disrupt these objects, and large parts of their stellar content can be expelled toward their surroundings. Hercules is likely in such a stage \citep{Roderick2015,Garling2018} and \citet{Kupper2017} recently modeled it as an exploding stellar system after two passages, concluding to an absence of a DM dominant component. This supports our conclusions, since Hercules lies on the relations shown in Figures~\ref{fig1} to ~\ref{fig3}.\\

\section{Discussion: the DM-free MW dSphs in the near-field cosmology context}

We propose a reasonable scenario that successfully explains the MW dSph properties and that excludes DM as being their main mass component. However, before accepting such a major change of paradigm in near field cosmology, one has to verify whether (1) most MW dSphs can be at their first passage, (2) what are their progenitors, and (3) how this could be consistent with their star formation histories. 

\subsection{A first passage for most dSphs?}
A very massive MW (e.g., $\sim$ 2$\times 10^{12} M_{\odot}$) would lead to more circular orbits for the dSphs after considering their {\it Gaia} DR2 proper motions.  For example, \cite{Fritz2018} found that one-third of them have orbits with eccentricities larger than 0.66 for their heavy MW mass model (1.6$\times 10^{12} M_{\odot}$), a fraction that increases to two-thirds for their light mass model (0.8$\times 10^{12} M_{\odot}$, see also \citealt{Bovy2015}).  Hence for a less massive MW, typically smaller than, e.g.,  1.3$\times 10^{12} M_{\odot}$, most dSphs including the numerous ultra-faint dwarfs, would have large orbital eccentricities suggesting a first infall.  This is consistent with the fact that most of them lie in a vast polar structure (Pawlowski et al. 2014) that also includes the Magellanic Clouds, which are at their first passage.

{\it Gaia} DR2 has also considerably improved our knowledge of the MW mass distribution up to 20-50 kpc, by establishing a more accurate rotation curve \citep{Eilers2019,Mroz2019}, better constraints on the Globular Cluster motions \citep{Eadie2019} and on the estimates of the escape velocity \citep[see also \citealt{Grand2019}]{Deason2019}. Even if they are still using extrapolations of the mass profile to the outer halo, these studies provide MW masses ranging from 0.7 to 1.3$\times 10^{12} M_{\odot}$, i.e., allowing a scenario of first passage for most dSphs.\\

{\bf \subsection{What are the progenitors of DM-free dSphs?}}
Due to their proximity, MW dSphs are by far the tiniest galaxies for which detailed properties can be derived. Their progenitors are likely gas-rich galaxies, for which \citet{Lelli2016a} provided the most insightful report of their properties, including their rotation curve, their gas content, and the 3.6$\mu m$ photometry for deriving their stellar masses.  However, among the sample of 21 dSphs studied here, only Fornax possesses a sufficiently large mass (3$\times 10^{7} M_{\odot}$), to be compared to, e.g., galaxies belonging to the lowest stellar mass range of {\bf the \citet{Lelli2016a} study, all of which are dIrrs.}

Here we consider the nine dIrrs for which the distance is accurately determined (from the tip of the Red Giant Branch) and with $L_{\rm 3.6\mu m}$ smaller than $10^{8} L_{\odot}$, corresponding to a stellar mass of 5$\times 10^{7} M_{\odot}$ according to \citet{Lelli2016a}. Among them, two (UGC 07577 and CamB) show no evidence for an excess of mass from their rotation curves, while for four (WLM, D564-8, ESO0444-G084, and NGC 6789) DM seems mandatory to explain their rotational velocity within $R_{2.2}$  (= to 2.2 times the disk scalelength).  For the remaining three galaxies (UGC 0443, NGC 3741, and DDO154) the necessity for DM within the $R_{2.2}$ radius is rather marginal, especially when accounting for systematic errors due to the assumed inclination. We verify that the above results are unchanged by increasing the $L_{\rm 3.6\mu m}$ limit to 2$\times 10^{8} L_{\odot}$.

It seems quite puzzling at a glance that only DM-free dIrrs are reaching the MW halo, while others would have avoided such an infall. However:
\begin{itemize}
\item A shallow-distributed DM could be preferentially stripped lowering its fraction in the transformed galaxy  \citep{Kazantzidis2011,Lokas2016}, including perhaps in their central region if DM particles have radially biased orbits compared to that of stars. 
\item Alternatively, the MW dSph progenitors could have come within a group \citep{D'Onghia08} at first infall, for which, e.g., the brightest galaxies could be DM-free, which cannot be excluded from the small statistics discussed above. Progenitors might also be gas-rich, tidal dwarf galaxies that are devoid of DM as suggested by \cite{Metz2007}. For example, tidal dwarf galaxies have been used with success to reproduce the Magellanic Stream, the Leading Arm, and the disks of satellites surrounding both MW and M31 \citep{Fouquet2012,Hammer2013,Hammer2015,Yang2014,Wang2019}. 
\end{itemize}

The above hypotheses might appear too radical and one may need to investigate further the progenitor problem. For example, the identification of gas-rich progenitors of the MW dSphs is possibly far more complex than that provided by the above short analysis. In an infalling DM-free dIrr scenario, the fraction of stars that have escaped the initial galaxy may be significantly high for some orbits, possibly reaching values up to 90\%, and also depending on a detailed description of the gas removal by ram-pressure exerted by the MW halo gas.  Under these conditions even Sculptor ($M_{stellar}$= 5$\times 10^{6} M_{\odot}$) could have a progenitor within the lowest mass range of \citet{Lelli2016a} galaxies. More problematic is the fact that it becomes unclear to which radius DM has to be investigated in the progenitors, perhaps in a region more central than $R_{2.2}$.

Another limitation is due to the fact that stellar masses of field dIrrs have been determined through stellar population synthesis models, i.e., in a very different way than for star-resolved dSphs. Detailed studies of the star formation and chemical enrichment histories lead to $M_{\rm stellar}$/$L_{V}$ values in excess of 2.1 for Fornax, and up to 4.3 for Sculptor \citep{de Boer2012a,de Boer2012b}. Quoting \citet{Kroupa2013}, "the chemical evolution modeling of the Fornax dwarf-spheroidal satellite galaxy demonstrates that this system must have produced stars up to at most about 25 $M_{\odot}$ in agreement with the prediction of the integrated galactic initial mass function (IGIMF) theory given the low star formation rate $\sim$ 3$\times 10^{-3}$ $M_{\odot} yr^{-1}$ deduced for this system when it was forming stars in the past." Since \citet{de Boer2012a,de Boer2012b} assumed a \cite{Kroupa2001} IMF that is extending to large star masses (120 $M_{\odot}$), the stellar mass of Fornax is perhaps underestimated. This can also apply to the Fornax progenitor as well as to most dIrrs, i.e., their IMF could be rather bottom-heavy according to the IGIMF theory \citep{Kroupa2013}, which could lead to large values of $M_{\rm stellar}$/$L_{\rm 3.6\mu m}$. For the later, larger values than 0.5 are not excluded by the baryonic Tully Fisher relation according to \citet{Lelli2016b}, and, for example, a value of 1 would be consistent with an absence of DM in the $R_{2.2}$ radius of UGC 0443, NGC 3741, and DDO154.

Even if one may succeed in identifying the progenitor of Fornax (and perhaps of Sculptor), it leaves unknown the properties of the progenitors of the 19 remaining dSphs studied in this paper. There have been pioneering studies of very tiny dIrrs including the {\it SHIELD} \citep{McNichols2016}, the {\it LITTLE THINGS} \citep{Oh2015}, and the {\it VLA-ANGST} \citep{Ott2012} surveys, reaching galaxies with stellar masses from 3$\times 10^{4}$ to 5$\times 10^{6} M_{\odot}$, i.e., below the smallest dIrr mass values of \citet{Lelli2016a}. \citet{Lelli2016a} have excluded the LITTLE THINGS dIrrs because of irregularities in their rotation curves possibly due to noncircular events.  Some attempts (see \citealt{Oh2015}) have been done to estimate their rotation curves, using the titled ring method that assumes an infinitesimally thin HI layer (see \citealt{Bosma2017}). However, such low-mass objects have almost a spheroidal geometry \citep{Sanchez-Janssen2010} and a velocity amplitude similar to that of their dispersion, the latter being mostly associated to star formation and turbulence \citep{Stilp2013a}. This likely hampers a robust determination of their rotation curve and hence of their DM content. \citet{Stilp2013b} concluded that "rotation curves for the smallest dwarfs in our ({\it VLA-ANGST}) sample may be difficult to obtain due to their complex velocity fields"  (see also \citealt{Ball2018} for Coma P, \citealt{Oh2015} for DDO 210, \citealt{McNichols2016} for AGC 748778, and \citealt{Bernstein-Cooper2014} for LeoP). This does not mean that there is no DM in the tiniest dIrrs, though it points out the need for a significantly more robust (3D) modeling of individual objects to disentangle effects due to star formation from those due to gravitation.%To verify whether DM-free dSphs are consistent with dIrr properties in the field requires to extend the \citet{Lelli2016a} study by several orders of magnitude toward the low mass end of  gas-rich galaxies.

\subsection{Are DM-free MW dSphs consistent with the star formation history?}
The above scenario might also appear at odds with the star formation histories of dSphs, which are often dominated by a single, very old population \citep{Weisz2014}.  
However, star formation histories could be complex and difficult to interpret, especially if unknown amounts of initial gas have been removed by ram-pressure effects leading to a subsequent decrease of self-gravity and then to an expansion of stars within a spherical symmetry. For example, during the last few hundreds of megayears, the star formation history of Fornax \citep{de Boer2013} has been consistent with a recent gas removal by ram-pressure and then with the tidal shock scenario. While this also applies to Carina, Leo I, and perhaps to Leo II, the cases of Sculptor, Sextans, UMi and Draco \citep{Weisz2014} are perhaps more problematic.  For example, why did the star formation in the Sculptor progenitor stop about 5 Gyr ago \citep{de Boer2012a}? Only a full hydrodynamical simulation with a well determined orbital history for Sculptor could help us to verify this potential inconsistency. Recently, {\it FIRE2} simulations \citep{Garrison-Kimmel2019} have shown that gas-rich dwarfs with stellar mass ranging from $10^{6}$ to $10^{7} M_{\odot}$ have similar star formation histories to Sculptor. Beside this, most of dSphs are much fainter than classical dSphs and they show large orbital eccentricities that may be consistent with a first infall \citep{Simon2018}, which might be difficult to reconcile with a single old stellar population \citep{Weisz2014}.  Perhaps their progenitors were so tiny that these systems were unable to form stars for very long periods. Alternatively, one needs to verify whether or not the faintest dSphs may also contain an intermediate age population, and this is likely out of our reach with the present data because of the small number of giant stars in these very scarcely populated galaxies (V. Hill 2019, private communication).

%Tidal shocking is affecting dSphs in a very different way than tidal stripping. The latter 
%It has been argued that the previously modelled DM-free dSph progenitors by \cite{Kroupa1997,Klessen1998,Read2006} would have broad horizontal branches, due to the effect of a dominant tidal stripping.
 
%One may also wonder why are we, nowadays, witnessing such numerous events in the MW halo (Peebles, private communication)? This could be related to the infall of the Magellanic Clouds, which is already an exceptional event \cite{Robotham2012} that brings a considerable amount of gas into the MW \cite{Fox2014}. We also take note of the recent event in M31, which has been severely reshaped by a major merger, 2-3 billion year ago \cite{Hammer2018b}.\\

\section{Conclusion}

The DM content of dSphs is calculated through the DM projected density along the line of sight. %The later is proportional to the gravitational acceleration due to DM projected along the line of sight, which is aligned with the direction of the MW gravitational acceleration. % (the angle $\theta$ in Table 2 is used to correct the most nearby . 
Here we demonstrate that the recently improved data from \cite{Munoz2018} and \cite{Fritz2018} reveal that the gravitational acceleration attributed to the DM self-gravity is strongly anticorrelated to the MW distance. Assuming a mass modeling of the MW consistent with its kinematics and rotation curve, we find that the {\it los} gravitational acceleration attributed to the DM self-gravity is precisely equal to that caused by the MW tidal shocks on DM-free dSphs.
% The latter is justified for almost all dSphs but the two most distant, Leo I and II. The behaviour of these two galaxies that escape from both the impulse approximation and Eq.~\ref{MWtides} has to be studied after time integration since their properties could be related to accumulation of tidal stripping effects and reorganisation of their stellar content after a long passage into the MW halo that they are leaving as shown by their positive radial velocities \cite{Fritz2018}.\\

This leads us to conclude that all the observed properties of the dSphs can be explained by the action of the MW only, without having to include any DM in dSphs. In DM-dominated models of dSphs the DM is somewhat shielding the action of the MW, and  the DM total mass appears not to be correlated to the MW distance or its gravitational action. The last point is in contradiction with the observed anticorrelation ($\rho= 0.76$ for 21 dSphs, 0.83 without Leo I and Leo II) between the MW distance and the {\it los} projected acceleration, which is the only quantity related to observations. %{\bf This study based on } T
The probability that this is just a coincidence is only 3$\times 10^{-4}$, which can be conservatively considered as the chance that DM impacts the kinematics of dSphs.  %Instead of being due to DM, large values of dSph velocity dispersions are likely caused by MW tidal shocks exerted on DM-devoid dwarf galaxies. 
 This result questions the search of DM in these evanescent stellar systems, as well as our understanding of dwarfs in the Local Group and their role in cosmology. \\
To further investigate about the origin of DM-free MW dSphs would require (1) the detection and the analysis of their progenitors that could be extremely low-mass ($M_{\rm stellar}$ from $10^{4}$ to $10^{6} M_{\odot}$) gas-rich galaxies in the field, (2) a significant progress in estimating their stellar ages through the turn-over of their main-sequence stars, and (3) an accurate determination of dSph orbits and of the MW mass distribution to perform detailed and case-by-case hydrodynamical simulations.

% If you have acknowledgments, this puts in the proper section head.
\begin{acknowledgments}
We are very grateful to Hadi Rahmani for a careful reading of this paper. We warmly thank Pavel Kroupa for his very useful comments on an early version of the manuscript and Paolo Salucci for his suggestions. We are grateful to the referee for the very insightful comments that have considerably improved the paper's content. This work was granted access to the HPC resources of TGCC/CINES/IDRIS made by GENCI, and to MesoPSL financed by the {\it Region Ile de France} and the project Equip Meso of the program {\it Investissements d Avenir} supervised by the {\it Agence Nationale de la Recherche}. This work has been supported by the China-France International Associated Laboratory Origins. J.L.W. thanks the China Scholarship Council (No. 201604910336) for the financial support.
\end{acknowledgments}

\clearpage
\appendix
\section{The data}
\label{data}
\subsection{Selection of the dSphs}
\label{selection}
\citet{Fritz2018} provide one of the most complete and up-to-date list of 39 dSphs with spectroscopic data on individual stars that can be used to estimate their line-of-sight velocity dispersions ($\sigma_{\rm los}$). The list also includes objects whose nature is still under debate (e.g., Crater I). Our goal is to test the DM estimate in dSphs lying in the virial halo of the MW, i.e., within 300 kpc, leading us to exclude Phoenix (which is indeed not a dSph) and Eridanus II. Objects with only upper limits on $\sigma_{\rm los}$ must also to be rejected either because their nature as a dSph or a star cluster is under debate (Crater I) or because they cannot lead to DM mass estimates (Segue II, Hydra II, Triangulum II, Tucana III, and Grus I).  Velocity dispersion estimates are likely contaminated by the presence of binary stars, which particularly affects galaxies with only a few stars spectroscopically measured (see, e.g., \cite{Spencer2017}). In order to keep the number of dSphs sufficiently high we chose to keep targets having eight or more stars to measure $\sigma_{\rm los}$, i.e., excluding Bootes II, Carina III, Horologium, and Pisces II. Finally, since we aim at estimating the influence of the MW on the DM measurements, we chose to further exclude Carina II, Reticulum II, and Hydrus, which lie in between the MW and the LMC. 

It results that the sample of dSphs inhabiting the virial halo of the MW and having robust kinematics data for evaluating the DM estimates includes 24 galaxies, enabling us to study the correlation between their fundamental parameters, i.e., their luminosities, radii, velocities,and distances. 

\begin{figure*}
\includegraphics[width=6in]{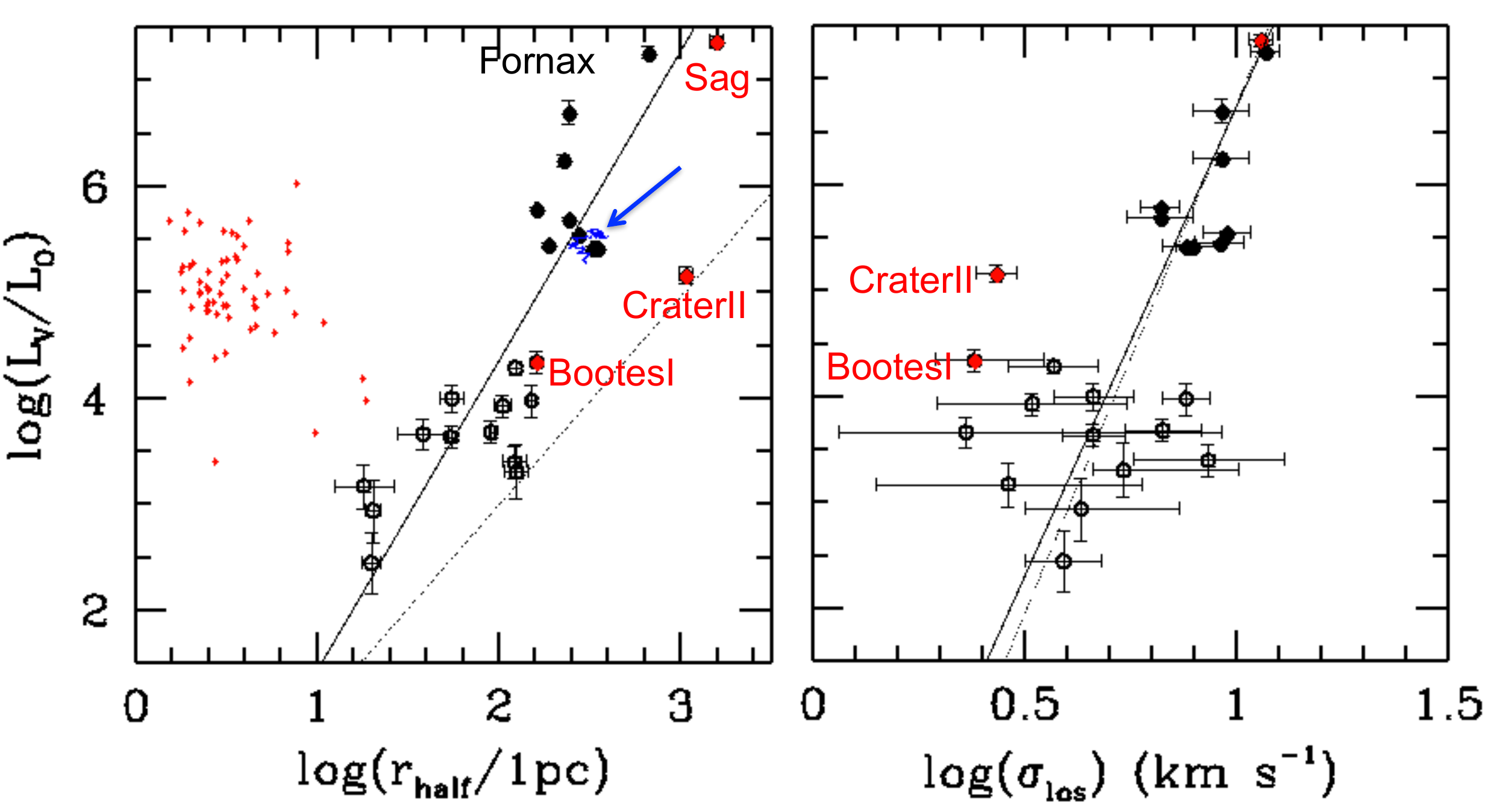}
\caption{Logarithmic correlation between visible luminosity ($L_V$) and half-light radius ($r_{\rm half}$, left panel) and line-of-sight velocity dispersion ($\sigma_{\rm los}$, right panel). Solid lines indicate the least-square fit of each relation, after excluding Crater II. In the left panel, small dots indicate the location of globular clusters from \cite{Moreno2014}, the dotted line shows a detection limit in surface brightness of 31 mag arcsec$^{-2}$, and the arrow indicates the track of the simulated dSph (same simulation than in Figures~\ref{fig3} and~\ref{movie}). In the right panel, the dotted line shows the correlation for the 21 dSphs of the sample after removing the three outliers (red points).}
\label{figA1}
\end{figure*}

\begin{figure}
\includegraphics[width=3.4in]{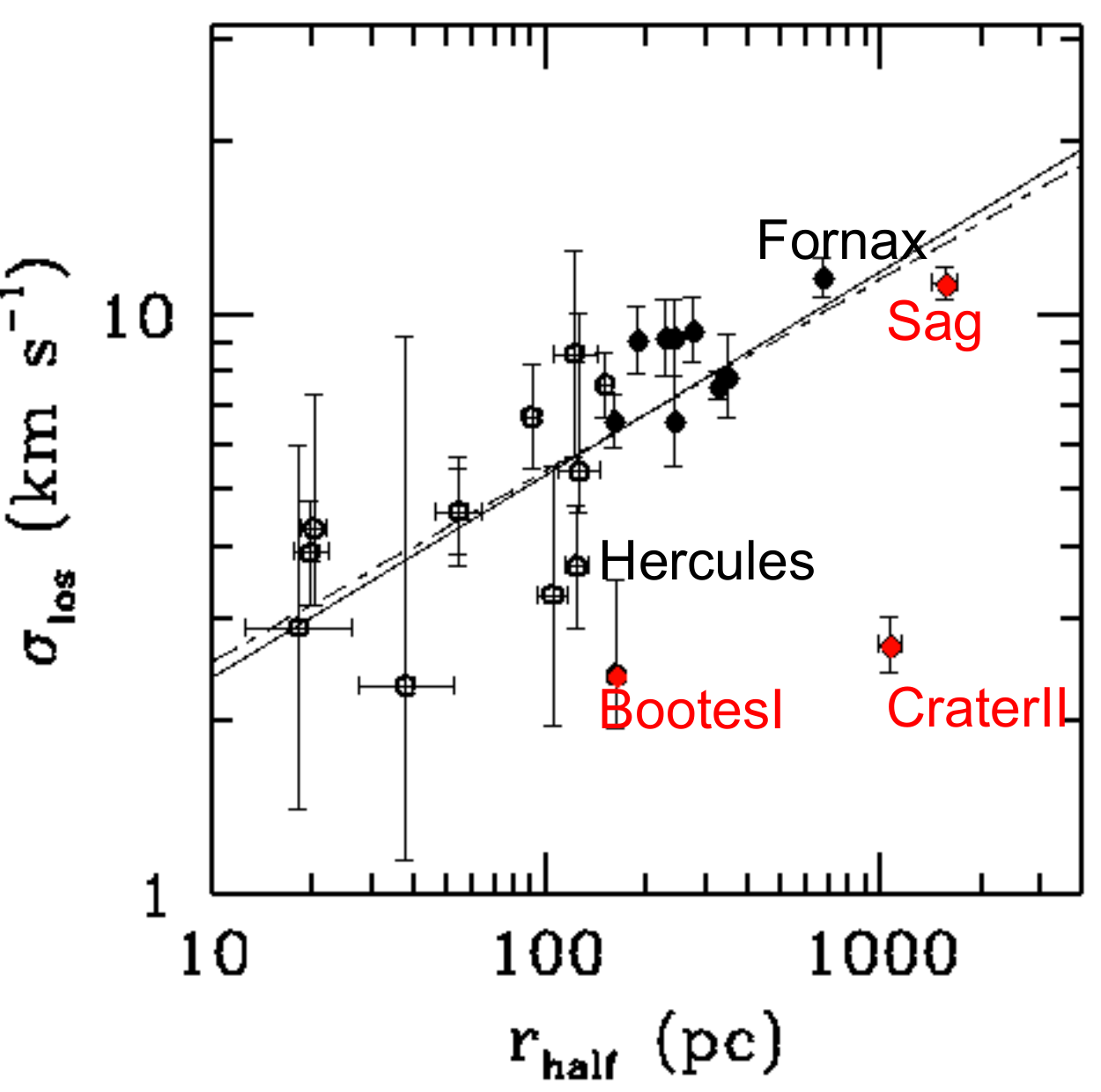}
\caption{Logarithmic correlation between half-light radius ($r_{\rm half}$) and line-of-sight velocity dispersion ($\sigma_{\rm los}$). Solid lines indicates the least-square fit, and the dashed-dotted line indicates the fit after combining the relations between ($L_V$, $r_{\rm half}$) and ($L_V$, $\sigma_{\rm los}$). The difference between the slopes calculated in both ways is taken as the uncertainty.}
\label{figA2}
\end{figure}

\subsection{Correlations between $L_{\rm V}$, $r_{\rm half}$, and $\sigma_{\rm los}$}
\label{correlations}
Table~\ref{tab1} lists the properties of the sample of 24 dSphs having robust kinematics data. Half-light radii ($r_{\rm half}$) and V luminosities ($L_{\rm V}$) are taken from \cite{Munoz2018}, assuming Plummer density profiles. Following \cite{McConnachie2012} we calculate the geometric mean half-light radius between the major and minor axes to account for the presence of highly elliptical systems. Left panel of Figure~\ref{figA1} shows the very strong correlation for the 24 dSphs but Crater II ($\rho$= 0.87, t = 8.1) between $L_V$ and $r_{\rm half}$, leading to  $\log(L_{\rm V})= 2.93 \log(r_{\rm half}) - 1.48$. Such a correlation can also be seen in Fig. 6 of \cite{McConnachie2012}. This means that the visible luminosity scales as $r^3_{\rm half}$. Surface-brightness limit may affect the scaling. However, only the discovery of many Crater II type galaxies could change it significantly.\\

 Assuming a Plummer profile, dSphs have an average central density of $\nu_0$= 0.0079 $L_{\odot} pc^{-3}$, a value about 20 times smaller that of the stellar density in the solar neighborhood. While the scatter between dSph values is quite large (the minimal value is $\nu_0$= 0.00025 for Aquarius, a value $\sim$ 350 times smaller than the largest one, for Leo I), the range of $L_V$ values is huge (a factor of 21,000). Also, note that Crater II is definitively an outlier with an ultra-low density, 33,000 lower than the average of other dSphs. 

Figure~\ref{figA1} (right) shows that $L_{\rm V}$ scales with $\sigma_{\rm los}^{8.88}$ ($\rho= 0.68, t=4.2$), i.e., with more than twice the power value expected from the baryon Tully Fisher relationship. This discrepancy has already been noted by \cite{McGaugh2010}. Combined together, the two relations displayed in Figure~\ref{figA1} lead to the relationship between $r_{\rm half}$, and $\sigma_{\rm los}$ shown in Figure~\ref{figA2} (see the dashed-small dashed line). The latter has also been found by \citet[see their Fig. 6]{Walker2009}, though the improved data measurements from \cite{Munoz2018}, as well as our choice of a geometrical mean for $r_{\rm half}$ (see, e.g., the Hercules location that is no more discrepant), have considerably reduced the scatter and increased the significance ($\rho$= 0.76, t = 5.4 for 23 galaxies). We also notice that the slope of the relation (0.37$\pm$0.04, see Eq.~\ref{sig_r}) is about twice that from \cite{Walker2009}.

%Central density $\nu_0$ does not correlate with and other parameters ($L_V$, $\sigma_{\rm los}$, $r_{\rm half}$ or $D_{\rm MW}$), suggesting it is intrinsic to each dSph and their formation history. Such a behaviour could have been interpreted as  $L_V$ representing the volume then the mass of all these systems, but Crater II. However, these dSphs inhabit the MW halo and if they have been supported only by their stellar mass, almost instantaneously they should have been tidally destroyed by the MW gravitation. 
\subsection{Correlations with $D_{\rm MW}$}
\label{correlationsDMW}

\begin{figure*}
\includegraphics[width=6.8in]{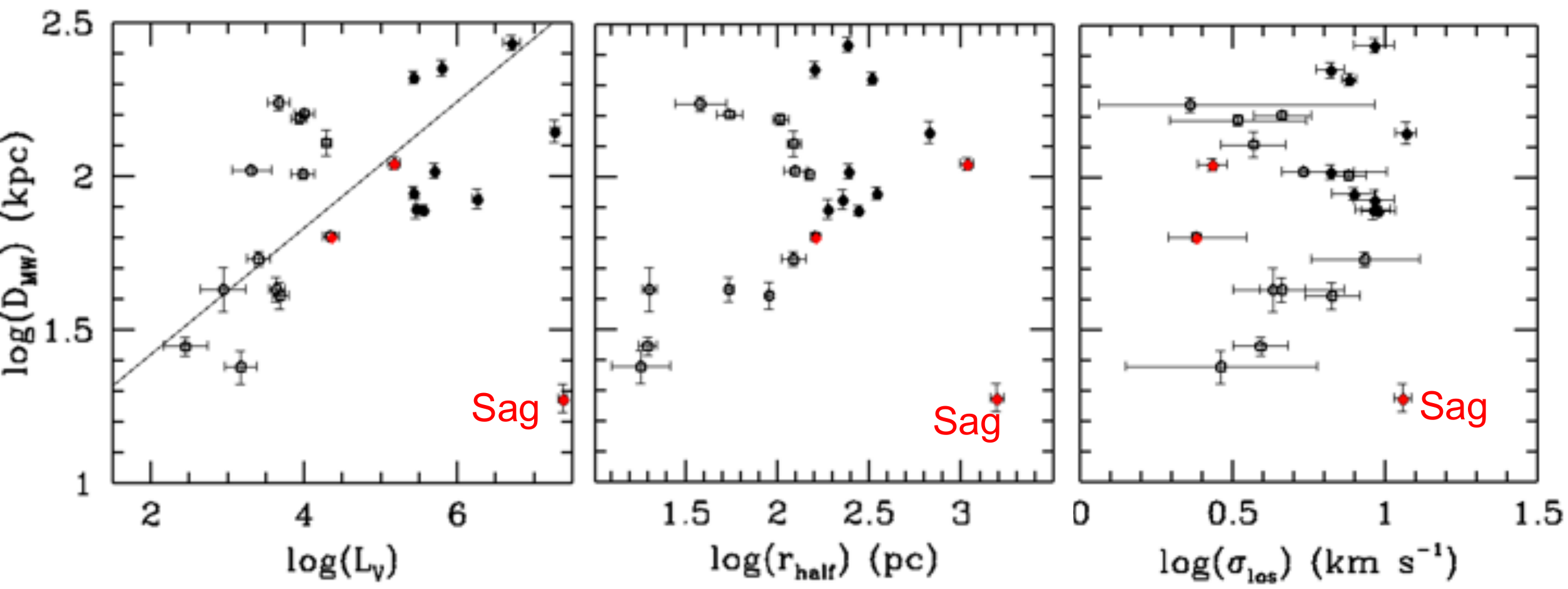}
\caption{{\it From left to right}: distance to the MW center against visible luminosity ($L_V$), half-light radius ($r_{\rm half}$), and line-of-sight velocity dispersion ($\sigma_{\rm los}$).}
\label{figA3}
\end{figure*}

Figure~\ref{figA3} shows the correlations between $L_{\rm V}$, $r_{\rm half}$, and $\sigma_{\rm los}$ with the MW distance. After discarding the outlier Sagittarius, there is a weak correlation between the $V$ luminosity and the distance ($\rho= 0.56, t=3.1$), which can be explained by the fact that only the brightest dSphs have been fully detected in the MW halo. The strength of the $L_V$-$r_{\rm half}$ correlation is sufficient to explain the very weak correlation between $r_{\rm half}$ and distance ($\rho= 0.42, t=2.2$). Note that $\sigma_{\rm los}$  does not correlate at all with distance, as it is also shown in Figure~\ref{fig2}.

It results that both Sagittarius and Crater II are outliers in the relations between the MW dSph fundamental properties. This is not surprising since the former was never considered for estimating the DM content as it is evidently dominated by MW tidal forces, which is further evidenced by the associated gigantic stream \citep{Ibata2003,Majewski2003}. Crater II was discovered recently (see \cite{Caldwell2017}) and was immediately considered as an outlier, due to its extremely small stellar density. From examination of Figures~\ref{figA1} to \ref{figA3} we find that while being less compelling, Bootes I is also an outlier when it is compared to the properties of the 21 remaining dSphs. Indeed Bootes I has two two distinct components with very discrepant velocity dispersions, and \citet{Simon2019} advised against using these data to derive Bootes I dynamical properties. The 21 remaining dSphs will be considered as the sample of ordinary dSph galaxies, and all Figures in the main manuscript are based on this sample. Table~\ref{tab2} gives the quantities calculated in the paper assuming the MW model of \cite{Sofue2015}.

\begin{figure}
\includegraphics[width=3.4in]{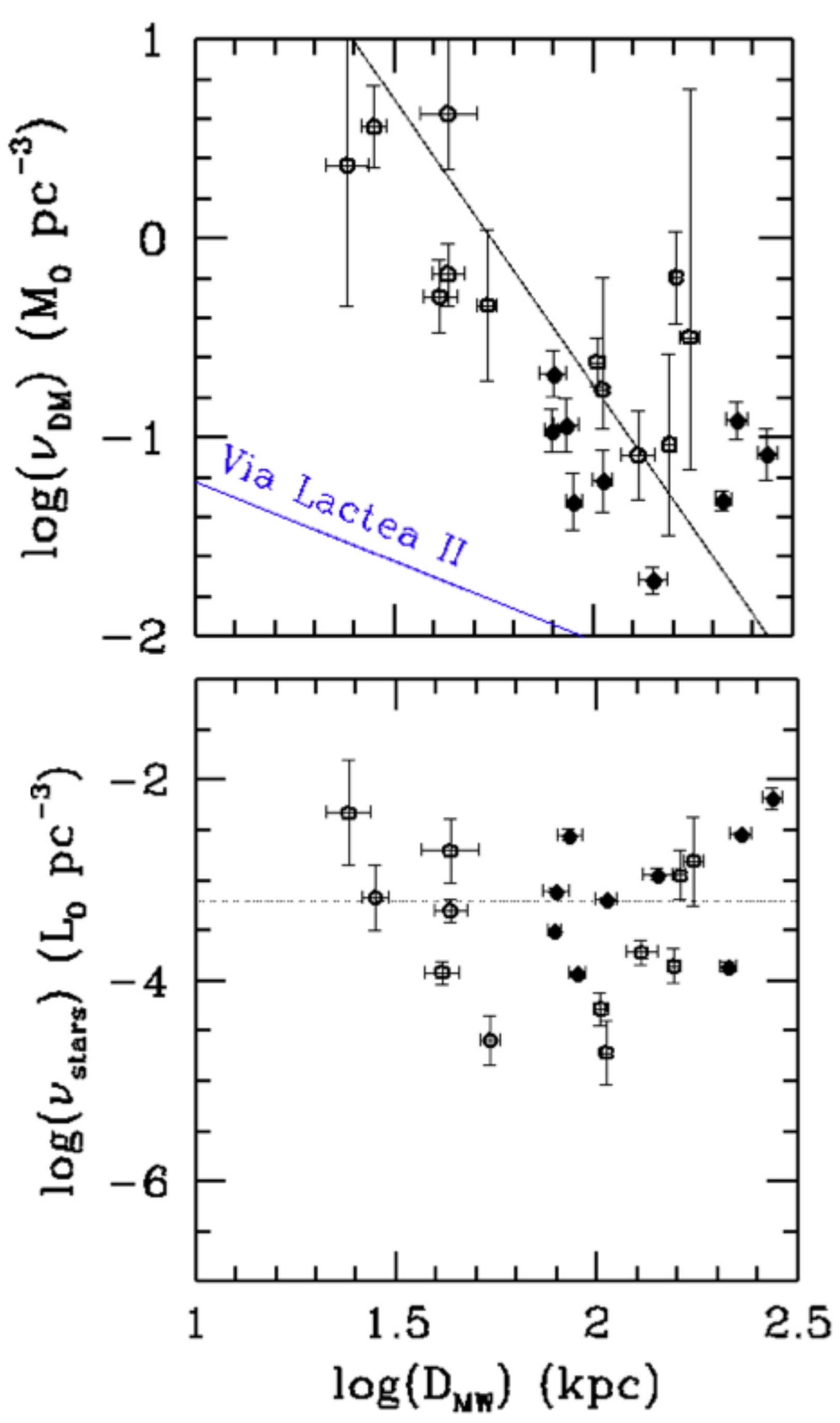}
\caption{Variations of dSph densities taken at r= $r_{1/2}$ with the MW distance for the sample of 21 ordinary MW dSphs. {\it Top}: DM density ($\nu_{\rm DM}$) calculated according to Eqs.~\ref{Wolf} and ~\ref{DMdensity}. It is logarithmically anticorrelated with  $D_{\rm MW}$, with $\rho$= -0.60, $t$= -3.3, and a logarithmic slope of -2.86 (full line) as indicated in the left panel of Figure~\ref{fig2}, and which contrasts with the prediction by the {\it Via Lactea II} simulation \citep{Diemand2008} as shown by the blue line. {\it Bottom}: star density ($\nu_{\rm stars}$) calculated at r= $r_{1/2}$ assuming a Plummer profile for dSphs \citep{Munoz2018}. It shows no correlation with $D_{\rm MW}$, with $\rho$= 0.15 and $t$= 0.66, the dotted line indicating the average value. }
\label{figA4}
\end{figure}

\section{Challenges to the scenario of DM-dominated MW dSphs}
\label{DMChallenge}

DSphs made of stars only are consistent with the fact that $L_{\rm V}$ ($\sim M_{\rm stellar}$) scales with $r_{\rm half}^3$ (Figure~\ref{figA1}). It is unexpected that tidal shocks affect the slope of this relation, because the energy provided by the MW tides $(1/2 \Delta \sigma^2_{MW})$ is approximately balanced by losses due to (too energetic) stars leaving the system \citep[see pages 435-436]{Binney1987}. In other words the reaction of the system to the MW tidal shocks would be to shrink and to loose stars, i.e., to decrease both $M_{\rm stellar}$ (or $L_{\rm V}$) and $r_{\rm half}$. Strong deviations from the baryon Tully Fisher (see Figure~\ref{figA1}) have led  \cite{McGaugh2010} to suggest a role for tidal effects. Combining $L_{\rm V}$ $\propto$ $\sigma_{\rm los}^{8.88}$ with the weaker correlation between $L_V$ and $D_{\rm MW}^5$, is consistent with a weak, or the absence of correlation of, between $\sigma_{\rm los}$ and $D_{\rm MW}$. This suggests that DM-free dSphs are consistent with all data and relationships presented in this paper.\\

%and tidal shocks would only expand quantities along the {\it los}. 
%It is also compliant with all relations shown in Figure~\ref{fig2}, and Eq.~\ref{MWtides} leads to $(\sigma_{\rm los}^2 - \sigma_{\rm stars}^2) / r_{\rm half}$ proportional to $D_{MW}^{-2}$. $L_V$ scales also with $D_{MW}^5$, and combined with  $L_V$ $\propto$ $r_{\rm half}^3$, it leads to $r_{\rm half}$ $\propto$ $D_{\rm MW}^{5/3}$, explaining why $\sigma_{\rm los}$, $M_J$ or $M_{DM}$ do not depend on $D_{MW}$. This suggests that DM-free dSphs are consistent with all the data and relationships presented in this paper.\\ 

Can these relationships be reproduced by models of DM-dominated MW dSphs? 
Pure DM subhalos in the MW potential \citep{Diemand2008, Moline2017} could yield an anticorrelation, such as that shown in the left panel of Figure~\ref{fig1}. This is because tidal stripping dominates in the diffusive regime (see \citealt{Binney2008}, P. 663) and leads to the  evaporation 
of particles having the lowest binding energy, which results in  
higher concentrations and densities \citep{Aguilar1993}. This particularly affects early infall,  and dSphs that inhabit the inner part of the halo.

Figure~\ref{figA4} shows that the DM density of dSphs is indeed anticorrelated with the MW distance, however, with a far much steeper slope (54 times higher) than that derived from Fig. 4 of \cite{Diemand2008}. %For the later, the sub-halo DM density increases by a factor 6.3 when $D_{MW}$ decreases from 250 to 30 kpc, while data indicates a factor 340 in the same range of $D_{MW}$. 
%Actually, the contrary would have been expected since outskirt regions should be more affected by tides and \citet{Diemand2008} calculated the density within a radius likely larger than $r_{1/2}$. 
 In these simulations, dSphs are infalling into the MW halo at different epochs. If tidal stripping caused the correlation shown in the top panel of Figure~\ref{figA4}, then
stars present at early epochs should be as affected by tides as DM, except if DM particles have much more radial motions than stars. According to \citet{Navarro2010} and \citet{Taylor2011}, DM particle motions may not be too anisotropic in the central regions of a DM halo. The bottom panel of Figure~\ref{figA4}
shows that on average the stellar density does not change with MW distance, which makes
unlikely an explanation based on DM tidal stripping, or, alternatively, that additional investigations are necessary, e.g, by considering the effect of baryons to the \cite{Diemand2008} prediction.\\
%Up to now we have been unable to figure out from the literature a standard scenario capable to face the challenges described in this study. Perhaps one has also to consider the large variety of astrophysical processes that have been considered in different manners in the standard scenario modeling e.g., 
It remains unclear how simulations (presence of gas in progenitors
and ram pressure stripping, stellar feedback and winds, radiative
 cooling, cosmic UV background affecting early star formation, see, e.g., \citealt{Kazantzidis2017}) of DM-dominated dSphs could reproduce the above relations, including Figures~\ref{fig1} and~\ref{fig2} and the scaling relations presented in the Appendices~\ref{correlations} and~\ref{correlationsDMW}. In spite of our efforts to investigate many simulations that have been performed to reproduce the Local Group content, we find none of them able to reach the properties of the lower half of dSph mass range, impacting their ability to reproduce the relationships discovered in this paper. The main limitation is the resolution, e.g., in one of the highest resolution {\it APOSTLE} simulations \citep{Fattahi2018}, the softening length is 134 pc leading to a spatial resolution of $\sim$ 400 pc, i.e., insufficient for most dSphs (see Table~\ref{tab1}). However, this limitation should soon be  overcome, e.g., by extreme resolution simulations \citep{Wheeler2018}. that might be used in the future to test our relations between host and dSph galaxies. For the moment, even \citet[see their Fig. 7]{Wheeler2018} have not fully reproduced the relation shown in the left panel of Figure~\ref{figA1}.

\section{Accounting for uncertainties in the analysis of the anticorrelation between acceleration and MW distance }
\label{stats}
The validity of the correlation analysis done in Sect.~\ref{Jeans_val} and shown in the left panel of Figure~\ref{fig1} may be questioned
since both axes (here called {\it x} and {\it y}) suffer from observational errors, which are both heteroschedastic and 
sometimes large. For each dSphs $i$, we note $\epsilon_{x,i}$ the uncertainties of the measurement on $x_i$, $\epsilon_{y,i}$ the uncertainties of the measurement on $y_i$ 
and $R_i = y_i - (ax_i+b)$ the residuals from the fitting process.\\

However, the observed dispersion on $y$ originates not only from the 
measurement errors, but also from some (unknown) intrinsic scatter
$\sigma_{\rm int}$, which increases the observed dispersion, 
$\sigma_{y,i}^2 = \sigma_{\rm int}^2 + \epsilon_{y,i}^2$.
To account for this, we checked the significance of our correlation 
analysis by two different methods. \\

First, we made a weighted regression \citep{Ripley1987}
accounting for errors in both axes, $\epsilon_{x,i}$ and $\sigma_{y,i}$. 
Initially starting with a null intrinsic scatter, 
we iterate twice the regression for the estimation of this intrinsic scatter:
we compute the residuals, and we estimate 
the variance of the intrinsic scatter with the variance of the residuals minus 
the average variances of the measurement errors, 
$\sigma_{\rm int}^2 \approx \sum R_i^2/(n-2)-\sum\epsilon_{y,i}^2/n$, 
as it is a consistent estimator of the intrinsic scatter
when $\epsilon_{x,i}$ can be neglected. 
We then found that the slope is significantly negative 
with a p-value = 3 $\times 10^{-4}$. 

Second we used an MCMC based Bayesian hierarchical model to fit the 
data, which derives the posterior distribution function from the data, and
considers the observation uncertainties and intrinsic
scatter simultaneously\footnote{see \url{https://github.com/sanjibs/bmcmc}}. 
The slope is $-0.54\pm 0.15$, i.e. with a significance similar to what has 
already been found.\\

These results show that the correlation found at Figure~\ref{fig1}
does not occur by chance only.

%A movie (see Figure~\ref{movie}) describes the overall process from the gas removal by ram-pressure to the expansion of the stars, and then tidal shocks exerted by the MW. It is based on \citet[see model TDG3-rp28 in their Table 3]{Yang2014} modeling, for which we have been able to decrease the gravitational force softening to 0.1 pc instead of 50 pc for the stellar type of particles (17 $M_{\odot}/particle$) in dSphs. Given the stellar density in dSphs, there should be no or very few stars that can be affected by the undesirable effects due to the softening.
%Star formation and  kinematics analyses are mandatory to evaluate the properties of potential gas-rich progenitors of MW dSphs, as well as better understand the properties of tidal dwarf 

%\clearpage
\pagestyle{empty}
\begin{longrotatetable}
\begin{table}
\begin{center}
\caption{Observed Parameters and Their Error Bars. Note. (1) dSph name; (2) and (3) V luminosity and its uncertainty; (4) and (5) Half-light radius and its uncertainty; (6), (7), and (8) {\it los} observed velocity dispersion and its uncertainty; (9) \& (10) Distance to the Sun and its uncertainty; (11) and (12) Local Standard of Rest velocity and its uncertainty; (13) Distance to the MW center; (14) Stellar mass to V luminosity ratio: 2.5 for dSphs without young or intermediate age stars, and 1.5 for Carina, Fornax, and Leo I (see, e.g., \citealt{Lelli2017} and references therein); (15) Galactic Standard of Rest velocity; (16) Angle between two lines  passing by the Galactic center and the Sun and crossing at the dSph; (17) $\epsilon$= 1-b/a, a and b being the major and minor axis of the dSph. }
\label{tab1}
{\scriptsize
\begin{tabular}{ccccccccccccccccc}
\hline\hline
name        & $L_{V}$        & $\delta L_{V}$& $r_{\rm half}$ & $\delta r_{\rm half}$ & $\sigma_{\rm los}$ & +$\delta \sigma$ & -$\delta \sigma$ & $D_{\odot}$ &$ \delta D_{\odot}$ & $v_{\rm LSR}$ & $\delta v_{\rm LSR}$ & D$_{\rm MW}$ & $M_{\rm stellar}/L_V$ & $v_{\rm GSR}$ & $\theta$ & $\epsilon$  \\
 & ($10^6$ $L_{\odot}$) &  & (pc) & & ($km$ $s^{-1}$) & & & (kpc) & &  ($km$ $s^{-1}$) & & (kpc) & ($M_{\odot}/L_{\odot}$) & ($km$ $s^{-1}$) & ($\deg$) &\\
 (1) & (2) & (3) &(4) & (5) & (6) & (7)& (8) & (9)&  (10) & (11)& (12) & (13) & (14) & (15) &(16) & (17)\\
\hline
Draco2      & 0.001497  & 7.14 $10^{-4}$& 20.73 & 7.636 & 2.9 & 2.1 & 2.1 & 24.46 & 3 & -335.0 & 1.6 & 24 & 2.5 & -175.5 & 24.0 & 0.24  \\
Segue1      & 2.811 $10^{-4}$& 1.888 $10^{-4}$& 24.11 & 2.79 & 3.9 & 0.8 & 0.8 & 24.96 & 2 & 203.6 & 0.9 & 28 & 2.5 & 112.6 & 17.0 & 0.33  \\
UMaII       & 0.004838  & 0.001157& 136.3 & 5.325 & 6.7 & 1.4 & 1.4 & 35.69 & 4 & -115.4 & 1.9 & 41 & 2.5 & -34.73 & 13.0 & 0.56  \\
Willman1    & 8.77 $10^{-4}$& 5.971 $10^{-4}$& 27.7 & 2.4 & 4.3 & 2.3 & 1.3 & 39.43 & 7 & -9.447 & 2.5 & 43 & 2.5 & 34.51 & 9.8 & 0.47  \\
Coma        & 0.004351  & 0.001001& 68.59 & 3.615 & 4.6 & 0.8 & 0.8 & 43.33 & 4 & 103.5 & 0.9 & 43 & 2.5 & 81.89 & 10.5 & 0.37  \\
TucanaII    & 0.002513  & 8.975 $10^{-4}$& 156.3 & 23.68 & 8.6 & 3.5 & 3.5 & 58.64 & 3 & -133.0 & 2.0 & 54 & 2.5 & -204.1 & 0.0 & 0.39  \\
Draco       & 0.2822    & 0.01299 & 222.4 & 2.079 & 9.1 & 1.2 & 1.2 & 79.83 & 6 & -276.3 & 0.1 & 79 & 2.5 & -95.74 & 6.4 & 0.29  \\
UMi         & 0.3516    & 0.01617 & 407.0 & 2.0 & 9.5 & 1.2 & 1.2 & 76.92 & 3 & -235.4 & 0.1 & 78 & 2.5 & -84.65 & 6.0 & 0.55  \\
Sculptor    & 1.782     & 0.2295  & 276.4 & 0.9872 & 9.2 & 1.4 & 1.4 & 85.67 & 6 & 103.2 & 0.1 & 85 & 2.5 & 78.37 & 5.6 & 0.33  \\
Sextans     & 0.2612    & 0.01443 & 412.1 & 2.993 & 7.9 & 1.3 & 1.3 & 86.65 & 4 & 217.3 & 0.1 & 89 & 2.5 & 71.69 & 5.0 & 0.3  \\
UMaI        & 0.009591  & 0.003353& 234.2 & 10.01 & 7.6 & 1.0 & 1.0 & 97.81 & 4 & -52.95 & 1.4 & 102 & 2.5 & -7.887 & 4.7 & 0.59  \\
Aquarius2   & 0.00203   & 0.0012  & 160.0 & 24.0 & 5.4 & 3.4 & 0.9 & 108.1 & 3 & -67.03 & 2.5 & 105 & 2.5 & 41.55 & 13.0 & 0.39  \\
Carina      & 0.4921    & 0.02264 & 303.1 & 2.952 & 6.6 & 1.2 & 1.2 & 104.0 & 6 & 207.2 & 0.1 & 105 & 2.5 & 6.571 & 2.7 & 0.36  \\
Hercules    & 0.01934   & 0.003025& 221.1 & 17.4 & 3.7 & 0.9 & 0.9 & 134.8 & 12 & 62.04 & 1.1 & 129 & 2.5 & 146.5 & 2.7 & 0.69  \\
Fornax      & 18.56     & 2.39    & 792.5 & 2.837 & 11.7 & 0.9 & 0.9 & 139.4 & 12 & 41.98 & 0.1 & 141 & 1.5 & -34.03 & 3.2 & 0.29  \\
LeoIV       & 0.008555  & 0.002047& 114.3 & 12.03 & 3.3 & 1.7 & 1.7 & 154.8 & 6 & 131.3 & 1.4 & 155 & 2.5 & 13.3 & 3.0 & 0.17  \\
CVenII      & 0.01009   & 0.002969& 70.83 & 11.22 & 4.6 & 1.0 & 1.0 & 160.8 & 4 & -120.8 & 1.2 & 161 & 2.5 & -95.23 & 3.0 & 0.4  \\
LeoV        & 0.004659  & 0.001544& 50.41 & 16.15 & 2.3 & 3.2 & 1.6 & 173.6 & 10 & 172.3 & 3.1 & 174 & 2.5 & 58.53 & 0.0 & 0.43  \\
CVenI       & 0.2651    & 0.01463 & 437.9 & 12.59 & 7.6 & 0.4 & 0.4 & 211.5 & 10 & 40.65 & 0.6 & 211 & 2.5 & 78.16 & 0.0 & 0.44  \\
LeoII       & 0.6242    & 0.02298 & 164.7 & 1.926 & 6.6 & 0.7 & 0.7 & 224.7 & 14 & 78.34 & 0.1 & 227 & 2.5 & 23.31 & 0.0 & 0.07  \\
LeoI        & 4.987     & 1.285   & 287.9 & 2.133 & 9.2 & 1.4 & 1.4 & 269.4 & 15 & 277.2 & 0.1 & 273 & 1.5 & 173.6 & 0.0 & 0.3  \\
Sagittarius  & 23.4      & 3.343   & 1636.0 & 52.78 & 11.4 & 0.7 & 0.7 & 26.93 & 2 & 150.1 & 2.0 & 19 & 1.5 & 170.9 & 7.1 & 0.64  \\
Bootes1     & 0.02212   & 0.005088& 192.5 & 5.039 & 2.4 & 0.9 & 0.5 & 67.27 & 2 & 109.5 & 2.1 & 64 & 2.5 & 107.0 & 7.3 & 0.3  \\
CraterII    & 0.1481    & 0.02677 & 1066 & 86 & 2.7 & 0.3 & 0.3 & 112.6 & 5 & 85.33 & 99.0 & 111 & 2.5 & -73.96 & 4.2 & 0.0  \\
%Bootes2     & 0.001225  & 8.344 $10^{-4}$& 37.92 & 4.997 & 10.5 & 7.4 & 7.4 & 42.63 & 1 & -106.7 & 5.2 & 39 & 2.5 & -115.4 & 11.5 & 0.25  \\
%Horologium  & 0.002458  & 0.001315& 37.68 & 7.327 & 4.9 & 2.8 & 0.9 & 82.77 & 5 & 98.96 & 1.0 & 83 & 2.5 & -28.81 & 0.0 & 0.27  \\
%Pisces2     & 0.004211  & 0.001414& 64.32 & 10.05 & 5.4 & 3.6 & 2.4 & 183.2 & 2 & -222.2 & 2.7 & 182 & 2.5 & -75.1 & 0.0 & 0.4  \\
%CarinaII    & 0.00745   & 6.898 $10^{-4}$& 106.9 & 9.397 & 3.4 & 1.0 & 1.0 & 37.87 & 99 & 463.4 & 99.0 & 37 & 2.5 & 253.1 & 0.0 & 0.34  \\
%ReticulumII & 0.003227  & 0.001128& 49.71 & 1.753 & 3.3 & 0.7 & 0.7 & 33.64 & 99 & 50.41 & 99.0 & 33 & 2.5 & -91.48 & 0.0 & 0.58  \\
%Segue2      & 8.937 $10^{-4}$& 1.767 $10^{-4}$& 35.68 & 3.058 & 2.6 & 2.6 & 2.6 & 36.93 & 2 & -46.29 & 2.5 & 42 & 2.5 & 41.84 & 8.7 & 0.15  \\
%EridanusII  & 0.06026   & 0.004989& 188.3 & 18.06 & 6.9 & 1.0 & 1.0 & 363.3 & 99 & 60.41 & 99.0 & 365 & 2.5 & -67.69 & 0.0 & 0.35  \\
%Phoenix     & 0.9174    & 0.2039  & 458.4 & 101.0 & 9.3 & 0.7 & 0.7 & 419.2 & 19 & -24.01 & 1.0 & 419 & 1.0 & -103.2 & 0.0 & 0.4  \\
%Pegasus3    & 9.53 $10^{-4}$& 5.0 $10^{-4}$& 53.0 & 14.0 & 5.4 & 3.0 & 2.5 & 207.2 & 20 & -216.3 & 2.6 & 205 & 2.5 & -62.43 & 0.0 & 0.38  \\

\hline
\end{tabular}
}
\end{center}
\end{table}
\end{longrotatetable}
\pagestyle{plain}
%\clearpage

%\clearpage
\pagestyle{empty}
\begin{longrotatetable}

\begin{table}
\begin{center}
\caption{Quantities calculated in the paper assuming the MW model of \citet{Sofue2015}. Note. (1)  dSph name; (2) Twice the kinetic energy associated to the velocity dispersion of a Plummer body made of stars only (see footnote 3); (3) Same for the excess of velocity dispersion due to the MW tidal shocks (see Eq.~\ref{MWtides}); (4) $\Delta D_{MW}$/$D_{MW}$: relative variation of $D_{MW}$ that would provoke a change of $\pm$1/2 $\sigma_{starsonly}^2$ in the kinetic energy of a dSph dominated by MW tidal shocks; (5) $\alpha_{\rm MW}$ = 1 - ($\partial$$M_{\rm MW}$/$\partial$$D_{\rm MW}$)$\times$($D_{\rm MW}$/$M_{\rm MW}$); (6) Predicted velocity dispersion that is $\sqrt{\sigma_{stars}^2+\sigma_{MW}^2}$, in which $\sigma_{MW}$ is the dispersion caused by MW tidal shocks (see Eq.~\ref{MWtides}); (7) For comparison, the {\it los} observed velocity dispersion; (8) Crossing time approximated by $r_{half}/\sigma_{los}$; (9) Encountering time approximated by $\Delta (D_{MW})/v_{\rm GSR}$; (10) Dynamical time approximated by $r_{half}/\sigma_{\rm starsonly}$ (11) Orbital time approximated by $D_{MW}/(2\times v_{GSR}$); (12) \& (13) ratios of the encountering time to the crossing time, and to the virialization time (3 $\times t_{dyn}$), respectively.}
\label{tab2}
{\scriptsize
\begin{tabular}{ccccccccccccc}\\
\hline\hline
name  &   $\sigma_{starsonly}^2$ &  $\sigma_{MW}^2$ & $\frac{\Delta D_{\rm MW}}{D_{\rm MW}} $  & $\alpha$  & $\sigma_{pred}$    
                   &  $\sigma_{los}$  & $t_{cross}$  & $t_{enc}$ & $t_{dyn}$ &   t$_{orbit}$ & $\frac{t_{enc}}{t_{cross}}$   & $\frac{t_{enc}}{3\,t_{dyn}}$ \\
        & ($km^2 s^{-2}$) &  ($km^2 s^{-2}$) &  &-&   ($km s^{-1}$)      & ($km s^{-1}$)    &  (yr)&(yr) &(yr)& (yr) &   &       \\ 
 (1) & (2) & (3) &(4) & (5) & (6) & (7)& (8) & (9)&  (10) & (11)& (12) & (13)  \\
\hline
Draco2 & 0.09275 & 22.23 & 0.002086 & 0.5089 & 4.727 & 2.9 & 6.1 $10^{6}$ & 2.792 $10^{5}$ & 5.809 $10^{7}$ & 6.693 $10^{7}$ & 0.04577 & 0.001602\\
Segue1 & 0.01595 & 20.81 & 0.0003831 & 0.5021 & 4.564 & 3.9 & 4.953 $10^{6}$ & 9.325 $10^{4}$ & 1.53 $10^{8}$ & 1.217 $10^{8}$ & 0.01883 & 0.0002032\\
UMaII & 0.05991 & 56.2 & 0.0005331 & 0.5054 & 7.501 & 6.7 & 1.321 $10^{7}$ & 6.16 $10^{5}$ & 3.616 $10^{8}$ & 5.778 $10^{8}$ & 0.04663 & 0.0005679\\
Willman1 & 0.04869 & 11.98 & 0.002032 & 0.5074 & 3.47 & 4.3 & 4.591 $10^{6}$ & 2.478 $10^{6}$ & 8.946 $10^{7}$ & 6.098 $10^{8}$ & 0.5399 & 0.009235\\
Coma & 0.08948 & 32.2 & 0.001389 & 0.5074 & 5.684 & 4.6 & 1.158 $10^{7}$ & 7.141 $10^{5}$ & 1.781 $10^{8}$ & 2.57 $10^{8}$ & 0.06164 & 0.001336\\
TucanaII & 0.02305 & 54.32 & 0.0002121 & 0.521 & 7.372 & 8.6 & 1.389 $10^{7}$ & 5.494 $10^{4}$ & 7.871 $10^{8}$ & 1.295 $10^{8}$ & 0.003954 & 2.327e-05\\
Draco & 1.686 & 48.71 & 0.01731 & 0.5529 & 7.123 & 9.1 & 2.016 $10^{7}$ & 1.398 $10^{7}$ & 1.413 $10^{8}$ & 4.038 $10^{8}$ & 0.6934 & 0.03298\\
UMi & 1.442 & 72.35 & 0.009965 & 0.5517 & 8.607 & 9.5 & 2.813 $10^{7}$ & 8.988 $10^{6}$ & 2.226 $10^{8}$ & 4.51 $10^{8}$ & 0.3195 & 0.01346\\
Sculptor & 8.819 & 53.29 & 0.08274 & 0.5598 & 7.992 & 9.2 & 2.407 $10^{7}$ & 8.784 $10^{7}$ & 7.457 $10^{7}$ & 5.308 $10^{8}$ & 3.649 & 0.3926\\
Sextans & 0.8482 & 76.33 & 0.005556 & 0.5642 & 8.795 & 7.9 & 4.272 $10^{7}$ & 6.752 $10^{6}$ & 3.665 $10^{8}$ & 6.076 $10^{8}$ & 0.158 & 0.006142\\
 UMaI & 0.07161 & 27.45 & 0.001304 & 0.5775 & 5.248 & 7.6 & 1.931 $10^{7}$ & 1.651 $10^{7}$ & 5.485 $10^{8}$ & 6.33 $10^{9}$ & 0.8548 & 0.01003\\
Aquarius2 & 0.01819 & 20.99 & 0.0004332 & 0.5803 & 4.584 & 5.4 & 2.265 $10^{7}$ & 1.072 $10^{6}$ & 9.07 $10^{8}$ & 1.237 $10^{9}$ & 0.04731 & 0.0003938\\
Carina & 2.272 & 42.81 & 0.02654 & 0.5803 & 6.748 & 6.6 & 3.596 $10^{7}$ & 4.151 $10^{8}$ & 1.575 $10^{8}$ & 7.821 $10^{9}$ & 11.54 & 0.8788\\
Hercules & 0.1759 & 16.21 & 0.005425 & 0.6006 & 4.053 & 3.7 & 3.257 $10^{7}$ & 4.676 $10^{6}$ & 2.873 $10^{8}$ & 4.31 $10^{8}$ & 0.1436 & 0.005424\\
Fornax & 18.67 & 77.28 & 0.1208 & 0.6094 & 9.985 & 11.7 & 5.587 $10^{7}$ & 4.899 $10^{8}$ & 1.513 $10^{8}$ & 2.028 $10^{9}$ & 8.769 & 1.08\\
LeoIV & 0.09199 & 10.5 & 0.004379 & 0.6185 & 3.258 & 3.3 & 3.089 $10^{7}$ & 4.995 $10^{7}$ & 3.361 $10^{8}$ & 5.704 $10^{9}$ & 1.617 & 0.04954\\
CVenII & 0.2059 & 5.235 & 0.01967 & 0.6222 & 2.341 & 4.6 & 1.167 $10^{7}$ & 3.255 $10^{7}$ & 1.184 $10^{8}$ & 8.274 $10^{8}$ & 2.788 & 0.09167\\
LeoV & 0.1371 & 3.247 & 0.0211 & 0.6295 & 1.847 & 2.3 & 1.62 $10^{7}$ & 6.141 $10^{7}$ & 1.006 $10^{8}$ & 1.455 $10^{9}$ & 3.791 & 0.2034\\
CVenI & 0.9058 & 20.96 & 0.0216 & 0.6474 & 4.696 & 7.6 & 4.221 $10^{7}$ & 5.709 $10^{7}$ & 3.37 $10^{8}$ & 1.321 $10^{9}$ & 1.353 & 0.05646\\
LeoII & 4.4 & 9.097 & 0.2419 & 0.6539 & 3.792 & 6.6 & 2.356 $10^{7}$ & 2.305 $10^{9}$ & 7.412 $10^{7}$ & 4.766 $10^{9}$ & 97.87 & 10.37\\
LeoI & 13.91 & 10.4 & 0.6687 & 0.6698 & 5.205 & 9.2 & 2.563 $10^{7}$ & 1.029 $10^{9}$ & 6.322 $10^{7}$ & 7.697 $10^{8}$ & 40.16 & 5.427\\
\hline
\end{tabular}
}
\end{center}
\end{table}
\end{longrotatetable}
\pagestyle{plain}
\clearpage

% Create the reference section using BibTeX:
%\bibliography{basename of .bib file}

%% This command is needed to show the entire author+affilation list when
%% the collaboration and author truncation commands are used.  It has to
%% go at the end of the manuscript.
%\allauthors

%% Include this line if you are using the \added, \replaced, \deleted
%% commands to see a summary list of all changes at the end of the article.
%\listofchanges

\clearpage

\end{document}